%% file: main.tex
\documentclass[sigconf]{acmart}

\AtBeginDocument{%
  \providecommand\BibTeX{{%
    \normalfont B\kern-0.5em{\scshape i\kern-0.25em b}\kern-0.8em\TeX}}}

\copyrightyear{2024}
\acmYear{2024}
\setcopyright{rightsretained}
\acmConference[CHIWORK '24]{Proceedings of the 3rd Annual Meeting of the Symposium on Human-Computer Interaction for Work}{June 25--27, 2024}{Newcastle upon Tyne, United Kingdom}
\acmBooktitle{Proceedings of the 3rd Annual Meeting of the Symposium on Human-Computer Interaction for Work (CHIWORK '24), June 25--27, 2024, Newcastle upon Tyne, United Kingdom}\acmDOI{10.1145/3663384.3663389}
\acmISBN{979-8-4007-1017-9/24/06}

\usepackage[textsize=footnotesize]{todonotes}
\usepackage{xspace}

\begin{document}
\title[Understanding Generative AI-Assisted Data Analysis through Participatory Prompting]{``It's like a rubber duck that talks back'': Understanding Generative AI-Assisted Data Analysis Workflows through a Participatory Prompting Study}


\author{Ian Drosos}
\authornote{Equal contribution.}
\email{t-iandrosos@microsoft.com}
\affiliation{%
  \institution{Microsoft Research}
  \city{Cambridge}
  \country{UK}
}

\author{Advait Sarkar}
\email{advait@microsoft.com}
\authornotemark[1]
\affiliation{%
  \institution{Microsoft Research, University of Cambridge, University College London}
  \country{UK}}

\author{Xiaotong (Tone) Xu}
\email{xt@ucsd.edu}
\authornote{The author was affiliated with Microsoft Research when this research was conducted.}
\affiliation{%
  \institution{University of California San Diego}
  \city{La Jolla}
  \country{USA}}

\author{Carina Negreanu}
\email{cnegreanu@microsoft.com}
\affiliation{%
  \institution{Microsoft Research}
  \city{Cambridge}
  \country{UK}
}

\author{Sean Rintel}
\email{serintel@microsoft.com}
\affiliation{%
  \institution{Microsoft Research}
  \city{Cambridge}
  \country{UK}
}

\author{Lev Tankelevitch}
\email{lev.tankelevitch@microsoft.com}
\affiliation{%
  \institution{Microsoft Research}
  \city{Cambridge}
  \country{UK}
}

\renewcommand{\shortauthors}{Drosos and Sarkar, et al.}

\begin{abstract}Generative AI tools can help users with many tasks. One such task is data analysis, which is notoriously challenging for non-expert end-users due to its expertise requirements, and where AI holds much potential, such as finding relevant data sources, proposing analysis strategies, and writing analysis code. To understand how data analysis workflows can be assisted or impaired by generative AI, we conducted a study (n=15) using Bing Chat via participatory prompting. Participatory prompting is a recently developed methodology in which users and researchers reflect together on tasks through co-engagement with generative AI. In this paper we demonstrate the value of the participatory prompting method. We found that generative AI benefits the information foraging and sensemaking loops of data analysis in specific ways, but also introduces its own barriers and challenges, arising from the difficulties of query formulation, specifying context, and verifying results. 
\end{abstract}

\ccsdesc[300]{Human-centered computing~HCI theory, concepts and models}
\ccsdesc[500]{Human-centered computing~Natural language interfaces}
\ccsdesc[100]{Computing methodologies~Natural language processing}
\ccsdesc[100]{Computing methodologies~Neural networks}
\ccsdesc[300]{Social and professional topics~User characteristics}
\ccsdesc[500]{Human-centered computing~Participatory design}
\ccsdesc[100]{Computing methodologies~Machine learning}
\keywords{}

\begin{teaserfigure}
    \centering
  \includegraphics[width=0.5\textwidth]{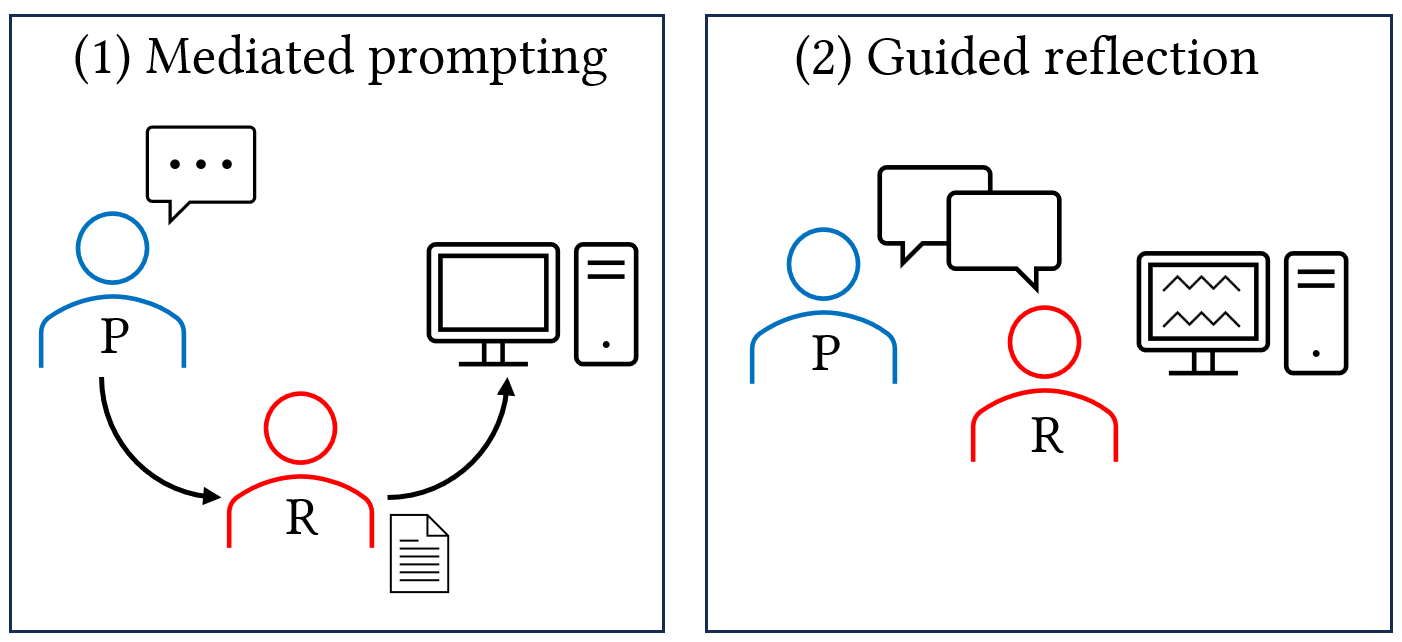}
  \caption{The turn-taking phase of the participatory prompting method. (1) Mediated prompting: the participant (P, blue) expresses their intent. The researcher (R, red) formulates a prompt based on this intent and a set of pre-prepared prompting strategies, and enters the prompt into the system. (2) The participant reflects on the result, guided by the researcher, and forms their next intent, after which the study returns to step (1) for the next turn.}
  \Description{A two panel image with diagrams showing the process of participatory prompting.}
  \label{fig:pp_turn_taking}
\end{teaserfigure}


\maketitle
\newcommand{\TOOL}{\textsc{TONETREETOOL}\xspace}

\section{Introduction}

End-user tools based on generative deep learning, i.e., ``generative AI'' (defined in Section~\ref{sec:genai-definition}) can substantially improve the ability of users to analyse and make sense of data, particularly those without formal expertise or training in data analysis. Data analysis workflows are notoriously tedious, challenging, error-prone, and have high expertise requirements. Generative AI significantly advances the state of the art in facilitating the authoring and debugging of data analysis scripts, reuse of analysis workflows, comprehension of analysis scripts, learning, and exploration \cite{sarkar2023eup_genai}. The potential change in user behaviour has been described as the \emph{generative shift} \cite{sarkar2023eup_genai}. The generative shift posits three axes of change: intensification (more sophisticated automation will be applied to existing workflows), extensification (more workflows will be automated), and acceleration (workflows which were previously costly will be applied in more contexts, as they become cheaper due to their automation).



An important user scenario for the generative shift is in \emph{end-user data-driven sensemaking}, that is, conducting analyses (often open-ended, ill-defined, and exploratory) within the context of some data (detailed in Section~\ref{sec:sensemaking-definition}). Classic examples of end-user data-driven sensemaking include personal and corporate budgeting, financial modelling in spreadsheets, and quantified self \cite{lupton2016quantified} activities. Less conspicuous examples include travel planning, or choosing a restaurant to visit or film to watch. These involve a mixture of qualitative and quantitative information, and of subjective and ``objective'' criteria; to choose a film, one might consider one's personal preferences and mood, the preferences of any companions, one's reactions to the trailer, critical reviews and ratings, film duration, genre, director, cast, and so on. 

As previously noted, generative AI has many applications in data-driven sensemaking. It can suggest relevant datasets or analysis procedures, write data transformation and analysis scripts or spreadsheet formulae, help debug or repurpose existing scripts, suggest subjective criteria for evaluating different options, teach the user how to apply an unfamiliar statistical procedure or tool, or even act as a critic or sounding board, to help the user decompose and refine an ill-defined problem. Faced with such a breadth of applications, the key question facing system designers is therefore one of scope: \emph{where are the greatest opportunities and challenges for improving the end-user experience of data-driven sensemaking with generative AI?} 




Our study is the first to apply the participatory prompting protocol by \citet{sarkar2023participatoryprompting} to explore the opportunities and challenges of generative AI for end-user sensemaking with data. 
Participatory promoting is a researcher-mediated interaction between the participant and a broad, open-ended AI system, such as OpenAI ChatGPT or Microsoft Bing Chat. The latter are ``broad'' in the sense that they are designed to support assistance in a wide range of workflows. 
By virtue of being researcher-mediated, participant experiences can be grounded in actual AI capabilities, scoped down by the researcher to a particular domain (in our case, data-driven sensemaking).  We further discuss the value of participatory prompting in the description of our method (Section~\ref{sec:method}).



Our study found that generative AI supports data analysis workflows in the information foraging loop by streamlining information gathering, and the sensemaking loop by helping users generate hypotheses and develop strategies to test them (Section~\ref{sec:genai-sensemaking}). However, we also found challenges to effective use of generative AI in data sensemaking workflows. These included forming effective queries, giving context to the AI, long or vague responses causing information overload, and frustrations with the verification of generated results (Section \ref{sec:sensemaking-barriers}). These results provide a range of implications for design, such as assisting users build detailed prompts that contain the context needed by AI to be effective, helping users verify AI responses, and better integration with feature-rich application workflows (detailed in Section~\ref{sec:design-implications}). 

As well as the domain-specific results, in this paper we also reflect on the value of the participatory prompting method for developing insights via mediated interaction that might otherwise remain unidentified. We discuss how it might expand to other fields of interest (Section \ref{sec:expandingMethod}), but also note some of its limitations in practice. These limitations include striking a balance in experimenter intervention to prevent over-influencing participant workflows, and potential inconsistencies between how researchers create and apply prompt strategies, which may reduce the reproducibility of results (detailed in Section~\ref{sec:limitations}).


\section{Background}
To clarify our guiding question, in this section we explain the concepts of sensemaking (Section~\ref{sec:sensemaking-definition}), generative AI (Section~\ref{sec:genai-definition}), and end-user programming (Section~\ref{sec:eup}), and summarise previous work on intelligent assistance for data analysis (Section~\ref{sec:analysis-assistance}). 


\subsection{Sensemaking}
\label{sec:sensemaking-definition}

We adopt Pirolli and Card's concept of sensemaking \cite{pirolli2005sensemaking}, which shares roots with Weick's~\cite{weick_socialpsychorgz_1969,weick_sensemaking_1995} organizational sense-making, but is focused on data analysis rather than social psychology. Sensemaking is the process by which individuals gather information, represent it schematically for interpretation, and develop insights into its meaning to create useful knowledge products. Sensemaking involves two iterative processes: (1) information foraging \cite{pirolli1999information} and (2) hypothesis development and testing (the latter by itself is also called the ``sensemaking loop''). 


The sensemaking framework is heavily influential and has been applied to understand data analyst workflows in multiple scenarios, such as navigating large datasets \cite{russell1993cost}, and understanding unfamiliar data visualisations \cite{lee2015people}. Notably, the latter study suggested that novices struggle to construct correct initial mental models (``frames'') to inform exploration, tending to persist with incorrect frames. To support sensemaking, the authors suggest that system designers should consider strategies like scaffolded introduction of visualizations or targeted annotation to aid formation of valid initial mental models.

A recent study explored how novice data analysts make sense of computational notebooks \cite{chattopadhyay2023make}. They developed an interface called Porpoise that groups code cells and adds structured labels to support these tasks (thus implementing the scaffolding and targeted annotation suggested by previous work). A counterbalanced user study with 24 practitioners found Porpoise facilitated comprehension and supported the building of mental models compared to default notebooks.

\subsection{Definition of generative AI}
\label{sec:genai-definition}

The term ``generative AI'' is extremely broad and encompasses many types of systems \cite{sarkar2023eup_genai}. The term can variously refer to core algorithms (e.g., the transformer architecture), specific instantiated models (e.g., GPT-4), or fully productized systems consisting of an ensemble of models plus additional components (e.g., ChatGPT).


To provide clarity around this term, Sarkar \cite{sarkar2023eup_genai} defines generative AI as \emph{``an end-user tool, applied to programming, whose technical implementation includes a generative model based on deep learning''}. The term ``end-user tool'' refers to tools that end-users directly interact with, not the underlying algorithms or models. The tool may consist of an ensemble of models, heuristics, engineered prompts, and interfaces. The definition is restricted to generative models based on contemporary deep learning techniques. Finally, the definition is restricted to the programming domain. Examples that fit this original definition include code completion tools leveraging large language models such as GitHub Copilot, and naturalistic language programming in spreadsheets using such models.



In this paper we adopt the ``end-user tool'' and ``technical implementation [...] based on deep learning'' aspects of the definition, but rather than programming, our domain of interest is sensemaking with data. Thus, we define generative AI as \emph{``an end-user tool, applied to sensemaking with data, whose technical implementation includes a generative model based on deep learning''}.


\subsection{End-user programming}
\label{sec:eup}

End user programming refers to programming primarily for personal use rather than public use, with the goal of supporting one's work or hobbies rather than developing commercial software. While end user programmers prioritize external goals over software quality, they face many software engineering challenges such as requirements elicitation, design, testing, debugging, and code reuse. Ko et al. provide a survey of the field \cite{ko2011state}. 

Much end-user programming research has focused on spreadsheets. Many techniques help with authoring spreadsheets, ranging from templating systems \cite{engels2005classsheets} to programming by example \cite{gulwani2011automating}. Testing methods like WYSIWYT (What You See Is What You Test) integrate white box testing into spreadsheet use \cite{rothermel1998you}. Debugging tools analyse formula dependencies or suggest fixes \cite{ferdowsi2023coldeco, williams2020understanding}. Other work focuses on developing higher level abstractions to facilitate reuse within spreadsheets, such as lambdas \cite{sarkar2022end}, sheet-defined functions \cite{jones2003user, mccutchen2020elastic}, and grid-based reuse \cite{joharizadeh2020gridlets}. Previous research has variously explored how spreadsheets are comprehended \cite{srinivasa2021spreadsheet}, learned and adopted \cite{sarkar2018spreadsheetlearning}, or structured \cite{chalhoub2022freedom}. Sensemaking theory has also been applied to end-user programming, for example, to explain and scaffold end-user debugging strategies \cite{grigoreanu2012end, horvath2022using}.

While many studies have investigated the potential of AI assistance for data analysis (which will be detailed in Section~\ref{sec:analysis-assistance}), a relatively smaller number have focused on the impact of generative AI more broadly on the activities of programming and end-user programming. Notably, no prior studies have investigated how generative AI tools can impact the data-driven sensemaking workflows of end-user programmers.




In a study exploring the emerging paradigm of artificial intelligence-assisted programming \cite{sarkar2022programmingai}, the authors observed shifts in the workflows of programmers, away from directly writing code and toward identifying suitable opportunities for AI aid, forming mental models of when AI support benefits workflows, and evaluating AI-generated output. The challenge for programmers transforms from writing code to activities such as judiciously \emph{``breaking down prompts at the `correct' level of detail,''} seen as an emerging core programmer competency. Other challenges involve constantly gauging whether any given scenario warrants AI involvement and debugging model outputs post-generation. Working with AI demands qualitatively different skill sets from programmers than previous workflows. More broadly, the theory  of \emph{``critical integration''} \cite{sarkar2023exploring}, i.e., the effortful and conscious evaluation, repair, and integration of AI output into a partially automated workflow, appears to be representative of how AI integration affects knowledge work.

An open question in end-user programming research is: to what extent people will still need to write code directly, if generative AI can do this for them from natural language prompts \cite{sarkar2023eup_genai}? As generative models advance, the author argues, they may facilitate a significant expansion in the scope and scale of end-user programming activities. However, this ``generative shift'' also raises questions about the continued relevance of traditional programming languages as an interface. In confronting these questions, the author proposes the focus of end-user programming research should transition from improving formal system usage to new questions around how to design for control and explanation, while mediating user intent through natural language.

\subsection{Intelligent assistance for data analysis}
\label{sec:analysis-assistance}

AI assistance for data analysis has long been studied under the paradigm of ``Intelligent Discovery Assistants'' (IDAs). Serban et al. provide an overview of IDAs \cite{serban2013survey}, which predate generative AI technologies and instead rely on AI planning and expert system techniques. Previous research has also considered the end-user activity of \emph{interactive analytical modelling}, i.e., building machine learning models as part of data analysis \cite{sarkar2014teach,sarkar2015interactive,sarkar2016visual}, and developed design principles for designing tools for non-experts \cite{sarkar2016phd}.


More recently, AI assistance has been studied in connection with exploratory data analysis and computational notebooks. \citet{gu2023analysts} investigate how data analysts from diverse technical backgrounds verify analyses generated by artificial intelligence (AI) systems, finding that analysts shift between procedure-oriented and data-oriented workflows. \citet{mcnutt2023design} conducted an interview study exploring the design space of AI code assistance in notebooks. Among other observations, analysts varied in their preferences in terms of the context provided to the AI system (full context or user-specified), and how assistance should be integrated into the workflow (e.g., in inline cells, in a sidebar, via pop-ups etc.). Chen et al. present WHATSNEXT, an interactive notebook environment that aims to facilitate exploratory data analysis with guidance and a low-code approach \cite{chen2023whatsnext}. The tool augments standard notebooks with insight-based recommendations for follow-up analysis questions or actions. Li et al. present EDAssistant, an interactive system that facilitates exploratory data analysis (EDA) in Jupyter notebooks through in-situ code search and recommendation \cite{li2023edassistant}. Wang et al. investigate how professional data scientists interact with a data science automation tool called AutoDS to complete an analysis task \cite{wang2021autods}. They observed that data scientists expressed more confidence in their manually-created models than models from AutoDS, even though AutoDS models performed better.


A particularly relevant study is \citet{gu2023data}, who explored analysts' responses to AI assistance that supports \textit{planning} of analyses. They first identified categories of suggestions that such a system could provide, including about data wrangling, conceptual model formulation, operationalisation of constructs, results interpretation, and others. In their Wizard-of-Oz setup, participants interacted with a JupyterLab notebook and received proactive analysis suggestions from a human wizard interacting with a LLM behind the scenes (the wizard was able to observe the notebook for context). Participants' generally valued planning assistance in the form of suggestions, but found them cognitively effortful to consider. Suggestions were helpful when accompanied by commented code, provided at an appropriate time in analysts' workflows, and when matching the analysts' statistical background, domain knowledge, and own analysis plan. However, in some cases, analysts became distracted by the suggestions or over-relied on them.

Researchers have also explored AI assistance from a sensemaking perspective, albeit theoretically, and not yet with empirical evidence from users. Wenskovitch et al. conceptualize how human-machine teams could facilitate AI-driven data sensemaking \cite{wenskovitch2021beyond}. The authors propose four roles that humans may assume in such teams: Explorer, Investigator, Teacher, and Judge. Similarly, Dorton and Hall propose a ``collaborative'' human-AI framework for sensemaking in intelligence analysis \cite{dorton2021collaborative}, notwithstanding critiques of the term ``human-AI collaboration'' and the collaboration metaphor for human-AI interaction more generally \cite{sarkar2023enough}.

In summary of the previous work:
\begin{itemize}
    \item Sensemaking theory gives us a framework for understanding the process of analysing datasets, particularly with open-ended or ill-defined questions. It decomposes the process into a set of interdependent loops of activity, and exposes opportunities for tool design. Sensemaking theory has been applied widely to visual analytics, intelligence analysis, and aspects of software development and end-user programming. However, the broader process of data analysis by non-expert end-users has not been studied with a sensemaking perspective.
    \item End-user programming research addresses the needs and challenges faced by people, typically non-programmers, writing programs for their own use. A particularly important site of end-user programming activity pertinent to data analysis is the spreadsheet. Numerous studies have elaborated the challenges that spreadsheet users face in learning and comprehending spreadsheets, and writing and debugging formulas. Sensemaking theory has been applied to study some aspects of end-user programming, but the potential impact of generative AI on the broader end-user activity of data analysis has not yet been studied.
    \item Intelligent assistance for data analysis has been explored in a number of ways, such as suggesting analysis paths and automatic experimentation. Many augmentations of computational notebooks, a common site for exploratory data analysis, have been proposed. Sensemaking theory has been considered in the context of AI assistance for data analysis, but prior explorations have been theoretical. Moreover, the efforts in this space have largely been directed towards expert data analysts.
\end{itemize}

Crucially, what is missing from previous literature is an understanding of the potential opportunities and challenges with applying generative AI to data-driven sensemaking workflows conducted by non-expert end-user programmers. This is the gap we aimed to fill. This research objective is incredibly broad; we cannot claim to have answered it definitively. However, our study has significantly advanced our understanding of the issue over previous work, and thrown light on new phenomena arising from the confluence of generative AI and end-user data analysis.

\section{Method}
\label{sec:method}

\subsection{Participatory prompting}
\label{sec:pp-definition}

At this stage in generative AI's development, exploratory research questions can be difficult to interrogate in ways that provide sufficient balance of ecological validity with both system access and researcher control. While generative AI systems with low usage barriers are available off-the-shelf, they can be difficult to focus on the task at hand without blockages, hallucinations, or other non-task-related issues that derail engagement. Alternatives are limited prototypes, mock-ups, or design fictions that can be too far removed from the actual capabilities of the technology, and lead to participant responses being based on an imagined caricature of AI conditioned by media narratives.

The participatory prompting method, first proposed by~\citet{sarkar2023participatoryprompting}, aims to bridge this gap. Participatory prompting is a user-centric research method for eliciting AI assistance opportunities. The method combines principles of contextual inquiry and participatory design \cite{schuler1993participatory}, in which researchers mediate participant interactions with a real generative AI system.


In a participatory prompting study, researchers first identify a domain problem and the relevant form of generative AI system. They experiment with different prompting formulations to elicit targeted responses, and then recruit participants who bring self-selected scenarios within the domain, and potentially also resources to be used. Researchers then conduct sessions in which participants work through their scenarios in multi-step turns (illustrated in~\autoref{fig:pp_turn_taking}). 




A key advantage of participatory prompting over low-fidelity prototyping and Wizard-of-Oz methods is that it grounds studies in ``actually existing AI'' \cite{siddarth2021ai} capabilities rather than simulations or speculative design probes. A benefit in comparison to experiments with fully functional prototypes is that it can leverage off-the-shelf AI systems with minimal engineering costs, and flexibly explore different use cases during a study, for which a functional prototype might be too constrained. 

Participatory prompting studies also have an advantage over some forms of purely observational studies, because by virtue of being researcher-mediated, participatory prompting can account for discrepancies in participants' \textit{a priori} prompting strategies, enabling participants to be appropriately challenged while not fixating on practical problems in generative AI usage that are not relevant to the research questions. 

Participatory prompting may involve various kinds of researcher mediation. The format used in this study is that of the researcher-as-relay. In this form, a participant poses an open-ended query to the researcher. The researcher reformulates the query using prompting strategies, and sends this prompt to the model. The participant reviews, reflects on, and builds upon the model's response to determine their next query, guided by the researcher. The `dialogue' with the system and the participants' reflections, together with optional quantitative measures of interactions such as response satisfaction, can then be analysed. Other formats could include researcher-as-guide, where the participant directly interacts with the AI system but discusses their thought processes with the researcher.

The interaction between the participant and the researcher creates valuable opportunities to elicit participant reasoning. First, the researcher can probe participant reasoning turn by turn (or sets of turns, as appropriate), to capture sequential expectations and responses. Second, when the researcher is involved in the translation of participant queries into prompts, participants may see and comment on the researcher's prompting strategies as reference point in comparison or contrast to what the participant might have done without guidance. While in some research methods this could be seen as influence or bias, in the participatory design context, this collaborative engagement on solving the problem of prompting reveals the differences and similarities between users' and technologists' assumptions, methods, and success criteria, and hence where either social or technical interventions or features are needed.


\subsection{Preparation}
The first step of the participatory prompting method is to choose a suitable functional generative AI system as a representative of AI capabilities more broadly. This involved careful evaluation of the possible alternatives. Four candidates were considered: OpenAI Playground, OpenAI ChatGPT, Google Bard, and Microsoft Bing Chat. We tested the systems by eliciting multi-stage guidance for data analysis through example queries, examining the quality and potential reception of each system's responses in a manner similar to a cognitive walkthrough \cite{lewis1997cognitive}.

We noted how particular design decisions in each system shaped and imposed limitations upon discourse. For instance, ChatGPT, Bing Chat, and Bard, as consumer products, incorporate ``guardrails'' against content considered inappropriate by the system developers, e.g., violent or sexual content. At the time of our study, Bing Chat restricted conversational exchanges to fifteen turns. In contrast, the OpenAI Playground allows more unrestrained exploration, and options for model and parameter selection. For our purposes, such constraints did not definitively preclude any options. However, the proprietary and opaque nature of commercial systems does restrict controllability, and this may render them unsuitable for some investigations.

At the time of our study, Bing Chat had the unique ability to source knowledge from the Web within replies. In a data-driven sensemaking activity, this can enhance suggestions at each problem-solving phase, such as by identifying relevant open datasets, and retrieving tutorials and recommendations for tool features (such as spreadsheet formulae). We found that information from the Web significantly improved the breadth and utility of the AI responses. This outweighed other limitations, and we therefore chose Bing Chat for our study.


The next step is preparing prompting strategies for the study. The challenge of developing reliable and effective prompting strategies to optimize large language models' performance has been comprehensively documented \cite{zamfirescu2023johnny}. Users, particularly non-experts directly engaging with AI systems, struggle to devise suitable prompts to elicit high-quality responses. To overcome this limitation, the participatory prompting protocol involves the mediation of an expert researcher with knowledge and practice of prompt design, to help users formulate suitable prompts. Besides this, the mediation also helps users rapidly iterate on queries, can help users focus on the relevant aspects of the interaction and avoid distraction from incidental elements of the user interface that are not relevant to the research questions, and eliminate variations in typing speed, as the researcher relays user queries to the system, rather than the user interacting with it directly.

For our study, three researchers individually experimented with developing prompting strategies for Bing Chat across four weeks, using a range of real data-driven sensemaking tasks drawn from their own personal or professional experience, including quantitative analysis of a poem text, choosing a bar to visit with colleagues, developing a spreadsheet for evaluating World of Warcraft game strategies, selecting a car for purchase, and choosing a plot type for a statistical report. These interaction logs and screenshots, successful and unsuccessful prompting strategies, error recovery methods, and other observations were catalogued in a shared repository.

Through this process, we identified that despite having the capability to do so, Bing Chat did not consistently use information from the Web, render tabular data visually as a table, or attribute its sources. We developed prompting strategies through which this behaviour could be reliably induced when needed. It often provided multiple options without further support to the user for choosing between them; we developed prompts (e.g., ``use information from the Web'', ``cite your sources'', and ``show result in a table'') to induce more such support when needed. 


At the end of the experimentation period, the researchers convened to negotiate and codify a list of prompting strategies and how they would be applied in different situations that might arise during the study. Despite having access to a thus carefully designed ``bank'' of prompting strategies, we found that in practice a lot of ad-hoc and in-situ adjustment was needed (discussed in Section~\ref{sec:limitations}).

\subsection{Pilot}
We conducted a pilot study with a convenience sample of 2 regular spreadsheet users. The pilots revealed that it can be difficult for participants to choose a suitable seed problem that is complex enough to require generative AI assistance but simple enough to describe concisely. To address this, more guiding questions were added to help participants during the problem elicitation phase. We also recommended that participants prepare a problem in advance of the study if possible. Terms such as ``data-driven decision-making'' were unclear to participants and had to be clarified.

We found that 5-6 turns could be completed in the allotted time (45–60 minutes), eliciting detailed qualitative insights despite the small number of turns. The turn-taking phase could be extended if needed. Reflecting on responses and choosing a next step was the most time-consuming and insightful aspect of each turn. This led to us changing the Bing Chat system mode from ``precise'' to ``creative'' (the exact nature of these modes is proprietary, but the salient aspect is that the latter is more verbose and the responses typically carry more information), to give more to reflect on and help guide next queries.

If early responses were generic or unhelpful, participants lost motivation. To counter this, advancement questions were added to the protocol to suggest ways forward, like rephrasing queries. Participants also tended to use short queries typical of web searches, which were more likely to result in generic responses; we added guidance to explain that longer, conversational queries were more effective.

We included steps for experimenters to more deeply understand participant expectations, including desired output types, to avoid multiple incremental prompts, which while useful to study, could slow down the progress of the task and thus impair the study of more complex interactions with the AI. Prompts also needed to refer to previous outputs to maintain consistency in the system responses; we updated the protocol to include this. While not initially part of the protocol, we noted that it was useful for participants to explore and verify outputs online, thus navigating temporarily away from the chat session. Finally, we revised the protocol so that participant speculations about helpful system capabilities could be immediately tested, and barriers to sensemaking were specifically elicited.

\subsection{Participatory prompting sessions}
\label{sec:PP-sessions}
We conducted a study with a fresh sample of spreadsheet users (N=15, 5 women, 0 non-binary, 10 men). Participants were recruited partly via email from a database of spreadsheet users who signalled interest to take part in research studies, and partly through a recruitment consultancy firm specialising in user research with African participants. Participation was voluntary, and all participants were free to withdraw from the study at any time without penalty and without having to cite a reason. All recruited participants were compensated with a USD \$50 (or local currency equivalent) gift voucher for an online retailer. Participants read and signed a consent form detailing the study format, data collection, and risks. The study method and data collection protocols were reviewed and approved by our institution's ethics review board.

\input{tables/tbl_participants}

Participants provided demographic information (Table~\ref{tbl:demographics}) relating to their experience with spreadsheets, programming, and generative AI via a survey. We directly use the spreadsheet and programming experience \cite{sarkar2020spreadsheet}, and generative AI experience \cite{sarkar2023participatoryprompting} survey items, and corresponding integer coding scheme, from previous work. Participants varied in spreadsheet usage (1 beginner, 7 experienced and basic usage, 7 experienced and advanced usage) and generative AI usage (3 never used, 1 casually use, 6 occasionally use, 5 regularly use), as well as programming experience (7 never programmed, 3 novices, 3 moderately experienced, 2 experts).
Participants resided in various locations (7 in Africa, 3 in Europe, and 5 in North America). 





The study sessions were conducted remotely using a Microsoft Teams video call, with the researcher handling the interaction with Bing Chat which was screen-shared with remote control to the participant, so that they could view and explore the results. 


Experimenters first elicited an example problem from the participant before entering the turn-taking phase as previously outlined. At each turn the participant was asked to read the response from Bing Chat and reflect aloud on the usefulness of the response and if anything was surprising, inspiring, or confusing. The experimenter would then ask the participant if they wanted to follow up with Bing Chat on the response, ask another question relating to the original problem, or pivot to a new problem they were interested in, thus proceeding to the next turn on the basis of the participant's response.

We tried to stay neutral, passive, and open-ended in terms of affecting the topic that the participant wanted to work on (and how to follow up in each turn). However, many participants found it hard to imagine what they would want to do with Bing Chat, and then how to follow up its responses. As such, we had to be active in eliciting their thought process and moving the study forward, by suggesting options to follow up and drawing their attention to certain aspects of the output.

The degree to which the researcher needed to intervene to reformulate the participants' query into a prompt varied depending on the context. At one extreme, the intervention was extremely minimal: a participant would dictate a query for the researcher to type verbatim. At the other extreme, when the participant found it challenging to articulate their need concisely, the researcher proceeded by writing a candidate query and asked the participant to confirm or disconfirm it, e.g. ``does this prompt capture what you wanted to ask the system to do?'' In between these two extremes, we would express their query directly, but suggest the addition of context (what columns existed in their spreadsheet), or append a prompt from our list of strategies (e.g. ``output a table'', or ``cite your sources''). The level of researcher intervention and the impact of query reformulation on our findings is a complex issue and we address the trade-offs in detail in Sections~\ref{sec:expandingMethod}~and~\ref{sec:limitations}.



Participants worked with experimenters through several turns between Bing Chat prompt and response until the task was achieved, or the allotted time was reached. Finally, participants gave further feedback on their experience through semi-structured interviews.


\subsection{Analysis method}
We transcribed the audio recordings of participant think-alouds and interviews. One researcher initially organised participant quotes through affinity mapping \cite{Beyer1997} into four broad categories: remarks about interaction with AI, remarks about workflows, remarks about barriers encountered, and remarks about specific features. The organisation was negotiated with a second researcher. This categorisation was not the final analysis, but a data management step to facilitate the final analysis.

The final analysis was a directed, negotiated coding between two researchers, with the aim of discovering emergent themes. We coded remarks relevant to the question of how AI assistance can support data sensemaking workflows according to the main categories of activities identified by sensemaking theory. We report our findings organised by these frameworks in Section~\ref{sec:genai-sensemaking}.  We coded remarks relevant to the question of how AI assistance can create barriers for data sensemaking workflows according to the \emph{iterative goal satisfaction} framework, described in Section~\ref{sec:sensemaking-barriers}, which also reports the results accordingly.

Our final analysis relied on the application of prior theoretical frameworks to supply the basis of code organisation, and thus differs from the more commonly applied inductive approach \cite{braun2006using}. We were not developing a reusable coding scheme and quantitative measures of inter-rater reliability are inappropriate here. Instead, in accordance with qualitative coding best practices, the two researchers iteratively discussed their interpretation of the findings and negotiated each disagreement until it was resolved \cite{mcdonald2019reliability}. Our analysis focused on identifying and characterising themes, rather than on quantifying the prevalence of each in our sample. As such, it is not helpful to be concerned with the precise participant counts associated with each identified theme, although this may be inferred from the list of participant IDs mentioned under each theme.






\section{Results}
\label{sec:results}

\subsection{Overview of tasks}

Recall that we did not design study tasks \emph{a priori} but rather developed them in a participatory manner with each participant at the start of the study using task elicitation questions. This resulted in a set of unique but highly ecologically valid tasks that were directly relevant to each participant. Participants explored a variety of data sensemaking tasks during the study, most related to their professional work, but also some personal workflows such as job searching or scheduling a pub crawl. 
The full list of participant tasks elicited is given in Table~\ref{tbl:tasksv2}. 




\input{tables/tbl_tasks_v2}

Each participant's task involved key sensemaking activities when seeking assistance with different aspects of data analysis. 
As part of the information foraging loop, participants often began by describing their data and its format (e.g., row and column descriptions), and their overall analysis goal. Some participants even began by requesting that Bing Chat generate or find example data (this tendency is corroborated by previous studies of analysts, which have found that analysis often begins in the absence of data \cite{muaruacsoiu2016clarifying}). For example, P10 requested that Bing Chat provide a list of potential career paths they could follow based on their skills and experience as a History PhD candidate. When P10 found a career path that was interesting to them (archivist), they continued by requesting the requirements for that career and example open positions that they could apply to. These interactions represented data filtering and searching within the information foraging loop.

As part of the sensemaking loop, most participants asked Bing Chat for help with formulating potential research questions (\emph{hypothesis generation}) or strategies for analysing their data (P1-5, 8, 9, 11-15), code or formulas for a specific analysis (P1, 2, 5, 8), or step-by-step instructions for applying Excel features such as filtering and visualizations (P4, 7) (\emph{hypothesis testing}). For example,  P9 first asked Bing Chat about data analysis strategies using Excel for data they collected in a survey. P9 then iterated with Bing Chat to generate potential research hypotheses and analysis plans for testing them.

\subsubsection{Example turn-taking sessions}
\label{sec:PP-turntaking}
An illustrative example of a complete turn-taking session (P1's) is described as follows. P1 wished to analyse a dataset about ``cooperative behaviour in literature'' they had collected. P1's first mediated query told Bing Chat they had a spreadsheet with data, where the \emph{``rows are the data for `tales' and columns contain the data for `cooperative behaviour' of a certain tale (e.g., `brother saved brother')}. The query indicated that they \emph{``need a way to code each cell according to different categories, explain how to use a spreadsheet for this with an example.''} Bing Chat's response confirmed its understanding, suggested options like using Excel VBA, thematic analysis, Google Sheets, and SPSS, and gave an example table and showed how thematic analysis might be applied to complete the task. 

P1's mediated follow-up response was to quote the fourth suggestion (use SPSS) and ask how this could be done in R with another example. Bing Chat's response explained how to use R in R Studio, and provided R code to complete the task and visualize the output. Each section of R code also contained a natural language description of the code. 



P1's final mediated query asked for the same code but with their specific categories in mind by asking \emph{``show me how to do it when the categories follow Hamilton's categories of biological cooperation''}. P1 was satisfied with the response, and this ended the turn-taking session for this task.

Another example, in brief: P6 was currently apartment hunting, so their first turn involved asking the Bing Chat to recommend ways to sort apartments based on the data they had previously collected. Subsequent turns involved recommending alternative apartments based on their criteria (by searching the Web). Finally, the participant requested Bing Chat to draft a letter to landlords to request extra information that Bing Chat recommended P6 collect, since it was missing from the spreadsheet they made.

\subsection{Data sensemaking workflow support}
\label{sec:genai-sensemaking}

Participants saw generative AI as a versatile tool that enabled various stages of data sensemaking. P11 saw generative AI as useful for \emph{``any part of a workflow''}, from \emph{``starting a new project''} to \emph{``preparing PowerPoint slides''} for presenting the project. Several participants thought generative AI supported their workflow by making their \emph{``work easier''} (P2, 4) by streamlining the search for \emph{``the desired result''} (P4), adding new perspectives on how to analyse their data (P3), and \emph{``scaffolding''} the solution to a task to \emph{``speed up the process''} of working with data (P1). P8, a business owner, believed generative AI would \emph{``save time''} and \emph{``greatly decrease cost''} for many of the tasks they needed to perform.



\begin{figure}[tb]
  \centering
  \includegraphics[width=\linewidth]{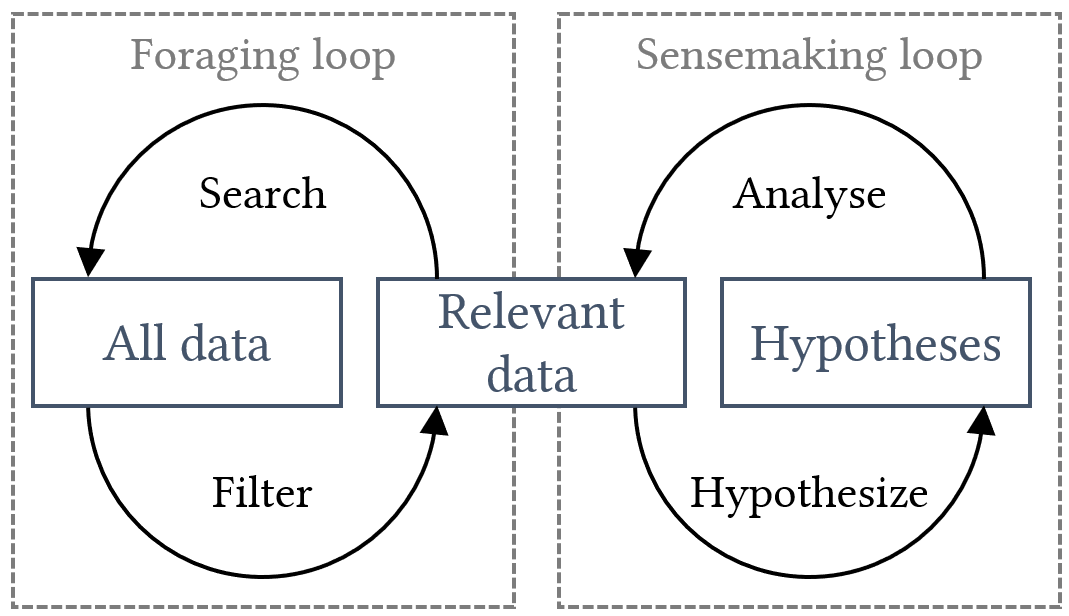}
  \caption{A simplified version of the data analyst's process, adapted from Pirolli and Card \cite{pirolli2005sensemaking}. The process consists of (1) a foraging loop, in which the analyst transitions back and forth between a large set of data and a smaller set of interest through searching and filtering, and (2) a sensemaking loop, in which the analyst transitions back and forth between relevant data and hypotheses through hypothesis generation and testing of the hypotheses (analysis).}
  \Description{A node link diagram. Boxes represent stages of a sensemaking workflow, with labelled arrows between them showing the transitions.}
  \label{fig:simple_sensemaking}
\end{figure}

The data analyst's work process as characterised by Pirolli and Card \cite{pirolli2005sensemaking} consists of two loops: an information foraging loop whose purpose is to identify a smaller set of relevant data out of a larger set, and a sensemaking loop whose purpose is to generate and test hypotheses. A high-level overview of this process is summarised in Figure~\ref{fig:simple_sensemaking}. This overview is helpful for further delineating aspects of our participants' experiences, presented in the following sections.

\subsubsection{Generative AI in the information foraging loop}

Several participants compared generative AI to traditional search. P9 thought that generative AI workflows improved upon the information overload caused by traditional search workflows since \emph{``search engines will give you multiple results, and it's very messy, but this [Bing Chat] directly gives one thing to do.''} P14 also enjoyed that generative AI output was specific to their question, while search results would require \emph{``converting the result''} to your specific task.

However, participants also cited concerns about the ability to apply generative AI tools to information foraging. P10 said their PhD research was \emph{``super-duper niche''} and frequently required them to \emph{``travel to archives all over the world''} to find data and thought generative AI would be unable to assist them for these types of tasks, because unlike textual data from the web, heterogenous archival data may not have uniform and easily accessible indices, and might be highly unstructured, mixed-media, only partially digitised, and therefore difficult or impossible for generative AI to operate over. 


\subsubsection{Generative AI in the sensemaking loop}
Participants noted the opportunity to be assisted both in generating hypotheses and in identifying strategies to test them. 
\label{sec:aiinthesensemakingloop}


\paragraph{Hypothesis generation} P4 thought generative AI was useful to \emph{``have another perspective, like conversing with another person to see how their perspective is different from yours''} which \emph{``could be inspiring''} for their own workflow by \emph{``giving a whole new way to do a task.''} Others believed generative AI was useful for brainstorming (P10) or getting unstuck by using the AI to give alternatives or options to explore (P7). P6 appreciated how Bing Chat's response considered aspects of the problem that they \emph{``didn't really get a chance to think about... so, it's good that Bing Chat was able to cover that as well.''} P13 said they were \emph{``directly inspired''} by Bing Chat, as it allowed them to \emph{``move further in the research analysis''} by introducing methods to do their task \emph{``in a different way''} than they had planned.

However, some participants were sceptical of using generative AI for their creative process (P1) or forming research questions (P9), and instead saw its primary application as being for specific data analysis tasks. This could be due to concerns about personal agency in the analysis process; for instance, P9 thought that even when generative AI generated useful text, it would still \emph{``miss your own style of writing''}.

\paragraph{Hypothesis testing} Participants noted how Bing Chat helped them avoid \emph{``spending ages try to figure out code''}(P1) and \emph{``insightful''} when offered analysis techniques they \emph{``had never thought of''} (P3). Participants also liked when Bing Chat provided \emph{``step-by-step process on how to get a chart in Excel''} that gave \emph{``headway on how to get the desired results''} (P4), and \emph{``some kind of direction''} (P5). P5 thought generative AI enabled this understanding by both \emph{``streamlining your thought process [...] with step-by-step instructions''} and giving \emph{``inspiration on how you can analyse data.''} P9 similarly valued the \emph{``step-by-step [instructions] on what to do''} and \emph{``other possible strategies''}. 


P12 likened generative AI to ``rubber duck'' debugging, an informal technique from software engineering where in order to fix a bug, the programmer explains their problem aloud to an inanimate object (archetypically, a rubber duck, hence the name) -- the idea is that verbalising the problem can often trigger the understanding and insight needed to fix the bug. P12 stated, \emph{``it's like a rubber duck that actually talks back and is useful.''}. This analogy highlights how, even if the AI system does not introduce new information, it may facilitate problem-solving and sensemaking by providing a channel for the reification and refinement of the user's thought process.

An additional benefit to the sensemaking loop was the ability to \emph{learn} new skills as part of the analysis process, which enriches the space of hypotheses it is possible to generate and test. These can be fairly straightforward technical skills, such as learning particular features of spreadsheet software. P7 had \emph{``a good learning experience''} in using an unfamiliar formula. P4 similarly \emph{``initially thought you could only create bar charts with a pivot table''}, but learnt from a Bing Chat suggestion that they \emph{``could just select the particular cell to create and insert the bar chart.''}

There is also the potential for learning broader skills. P5 saw Bing Chat's recommendations of unfamiliar functions and statistical packages as a potential \emph{``learning direction on how to go about carrying out descriptive statistics and visualizations to assist with that task.''} P10 saw generative AI as a potential learning surface that assists in critical thinking, because when P10 asked for a biography of Thomas Jefferson, the response did not initially raise the problematic issue of Jefferson's slave ownership, which P10 expected. P10 reflected that generative AI could be used to explore \emph{``what kind of questions we can ask and what kind of information is being omitted''}. This finding aligns with the constructivist theory of learning in interactive machine learning systems, which holds that users construct mental models of their task through iterative exposure to AI model responses \cite{sarkar2016constructivist}.







\subsection{Barriers to sensemaking with generative AI}
\label{sec:sensemaking-barriers}

Rather than thematising barriers according to the analyst process, we found that it is more helpful to consider them in terms of a workflow we term \emph{iterative goal satisfaction}. Broadly, this is the process by which a user satisfies a series of goals with AI assistance.



\begin{figure*}[tb]
  \centering
  \includegraphics[width=\linewidth]{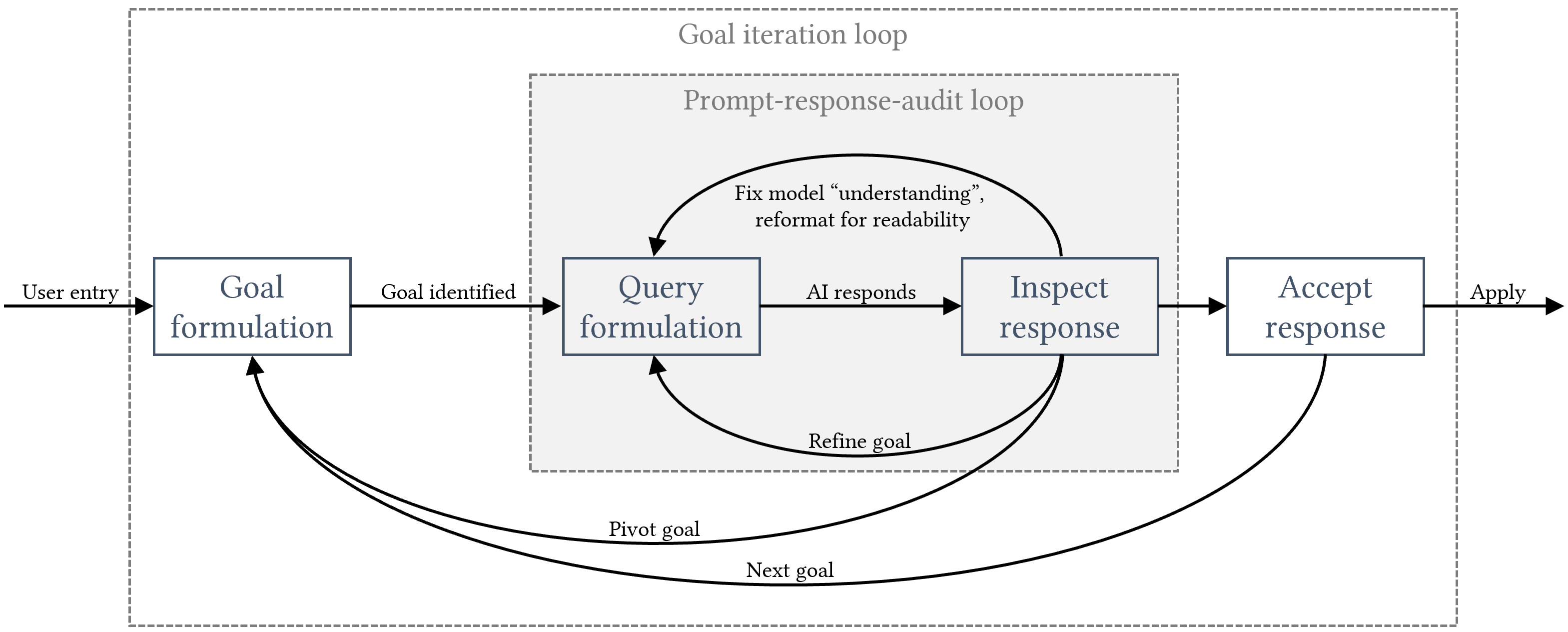}
  \caption{The workflow of iterative goal satisfaction with generative AI.}
  \Description{A node link diagram showing different phases of the user workflow with generative AI and labelled arrows showing transitions between them.}
  \label{fig:genAISenseModel}
\end{figure*}

The iterative goal satisfaction workflow is presented in Figure~\ref{fig:genAISenseModel}. The user moves through different phases: goal formulation, query formulation, and response inspection. There is an outer ``goal iteration'' loop as the user attempts to achieve a high-level goal, and an inner ``prompt-response-audit'' loop as the user attempts to achieve specific steps towards that goal. The elements of this workflow are as follows:

\begin{itemize}
    \item \emph{Goal formulation}: the user reflects on their goals, needs, intents, and research questions, and identifies a need for assistance where AI could be applied.

    \item \emph{Query formulation}: the participant composes the information, context, and data that the AI might need to address a goal (in our study, the query is relayed to the mediating researcher who then further shapes it into a prompt). Query formulation can proceed directly from goal formulation, or it may be in the context of iterating on a previously identified goal, as a result of having inspected a previous response (described next).

    \item \emph{Response inspection}: the participant checks for readability and relevance to the goal. If the output is readable and relevant, the participant reads with the aim of deeper comprehension, checking quality and correctness. If the response failed any checks, participants would either reformulate their query to attempt to elicit a better response, or change their overall goal. The sequence of query formulation and re-formulation in response to deficiencies identified by inspecting the output maps directly to the \emph{prompt-response-audit} cycle described by Gordon et al. \cite{gordon2023co}. 

    \item \emph{Response acceptance}: when the AI response satisfies their goal, participants might exit the goal iteration workflow entirely (e.g., to apply the results by copying a formula into a spreadsheet, or add code to their IDE), or develop a new goal. We thus observed two situations in which participants could develop entirely new goals: either as a result of having their previous goal satisfied, or a ``pivot'' as a result of inspecting a response and reflecting upon it. Consequently, we broaden Gordon et al.'s prompt-response-audit loop by showing that there are two distinct reasons for exiting it, and that it is itself part of a larger goal iteration loop.
\end{itemize}

With this picture of the iterative goal satisfaction workflow, we are in a better position to understand the barriers to effective sensemaking with generative AI encountered by our participants. Broadly, these fell into three categories: barriers to \emph{query formulation}, barriers to the \emph{utilisation of responses}, and barriers to \emph{verification and trust}. We detail each of these in turn.

\subsubsection{Barriers to query formulation}
Participants faced difficulties in understanding, gathering, and expressing their request. These are difficulties they experienced in their own articulation of their needs.

\paragraph{Detailed expression of intent} Part of the challenge was in fully articulating their need. Participants had trouble \emph{``wording it in the right way that the AI understands [...] writing [what is in your head] down is the hard part.''} (P1) and giving \emph{``a very explicit explanation in the prompt that is detailed''} (P13), though P1 noted that Bing Chat could generate helpful responses for \emph{``convoluted''} questions (i.e., prompts worded in a noticeably vague or unnatural manner). P5 similarly was frustrated by their inability to \emph{``really define the problem because there are a lot of components, a lot of things to factor in before clearly defining the problem.''}; it was challenging to \emph{``be as detailed as possible when you are putting information [into a prompt, but], you can't just be lazy about it and get the most useful answer [...] you have to feed [Bing Chat] with as much detail as possible.''} Such difficulties led to P9 asking \emph{``where should I learn this kind of stuff when I'm chatting with Bing Chat''}. 


Barriers to query formulation resulted in, but also stemmed from, inadequate output from the AI, with P12 stating \emph{``it is frustrating to figure out what is it that is being miscommunicated.''} P8 pointed out that \emph{``generative AI can't read your mind, so you just have to formulate your question `correctly'''}, and they would \emph{``be annoyed at myself for not writing the prompt correctly''} rather than blame the system for an inadequate output. Other participants similarly attributed this issue to their having \emph{``communicated `wrongly' at first''} (P4). P2 observed that the prompts that the experimenter wrote were \emph{``very different''} than their own in that they were more specific and \emph{``direct''}. P2 described their current prompting methods as \emph{``too general''} in comparison, and having difficulty understanding \emph{``where to start from''} when interacting with generative AI.
%

Participants developed strategies to manage the challenge of detailed expression. Participants used follow-up prompts to  \emph{``ask it specifically to focus''} on a specific part of their data (P3) or on a \emph{``specific list of categories''} (P4).  P1 thought the solution was \emph{``just asking the right questions''}, which meant being \emph{``clear and real specific in the details''}, though this was challenging and left them \emph{``a bit confused.''} P13's received a response localised to a different country, so they realised they should \emph{``be even more specific''} about their location. P5 decomposed their queries to \emph{``streamline them to focus on things I actually need and not just suggest the entire data analysis strategy.''} P12 thought they would improve their prompts by practising through \emph{``having to use it over and over again.''} Others developed more ad-hoc techniques, such as avoiding acronyms (e.g., `MSFT' instead of `Microsoft') (P6), to reduce the likelihood of miscommunication.

\paragraph{Determining and expressing context.} Participants were also challenged by the need to determine what contextual information was relevant to fulfilling or interpreting their request, and then articulating it. For example, after being recommended `thematic analysis' as a way to analyse their data (which was not applicable to the kind of data they had), P1 noted that giving context (in this case, information that would enable the system to rule out thematic analysis as a plausible method) to generative AI was important for making sure AI suggestions \emph{``actually work''} for the task and data. Participants drew a comparison to their experience of human-human collaboration. P12 found giving this context to generative AI was more difficult than giving it to a human co-worker, as they usually framed questions with what they had attempted previously and what went wrong before asking \emph{``what should I change?''} when asking a co-worker for help. P12 felt that their co-workers are \emph{``more familiar with examples''} that they would provide as context to their problem, and worried this context seemed more difficult to convey to the system.




Researcher mediation of prompts occasionally impacted participant awareness of these barriers. For example, researchers asked participants for needs around the data format of the response or gathered extra context about the problem being solved, which revealed to participants the specific prompting strategies we applied. Researcher-mediated queries served as a reference point for participants to compare their own experiences in forming queries.
While some aspects of effective prompting could be handled by the mediating influence of the researcher and thus ``smoothed over'' from the participant's perspective, as the examples above show, even with such guidance, participants are challenged by the activity of expressing their intent. 

\subsubsection{Barriers to utilisation of responses}
Participants faced barriers to being able to effectively use the responses, such as an overwhelming volume of information; poorly or incorrectly formatted results; output that while not strictly \emph{incorrect} was nonetheless incomplete or inadequate in some other qualitative manner; and responses which were not easily intelligible because they referred to unfamiliar concepts.

\paragraph{Volume of information in the response}

Bing Chat's responses were often lengthy, likely due to our choice of using Bing Chat's ``creative mode'' which is designed to be more verbose. This required the participant to read several paragraphs of text. P2 experienced information overload with the text results from Bing Chat, which they originally expected to be returned as a table or spreadsheet. P3 complained about the level of technical detail in one of the responses, finding it \emph{``not easily understandable for someone who is being introduced or does not have much experience in statistics''}. This points to the need for tailoring responses according to user expertise. When given different options to complete a task, P13 thought it was useful, but also \emph{``excessive information''}.

Excessive length also applied to generated code. P8 received \emph{``additional unnecessary code''} based on what they asked for, but nonetheless believed the result to be correct. In follow up, P8 asked Bing Chat multiple times to \emph{``make it [the code] shorter''}, until Bing Chat successfully reduced a 15 line function into 3 lines.

Participants' preferences regarding a suitable default length and contents for generative AI output varied (P3, 8, 10-12). For example, P3 preferred a specific order of generative AI output: first the answer, then an explanation of that answer, and finally an example of how to implement it in Excel. P8 shared a preference for seeing examples and expected Bing Chat responses go beyond \emph{``just some sort of summary''} by producing examples that apply Bing Chat's recommendations (e.g., showing how the A/B testing model Bing Chat generated might apply to a video advertising campaign for a company). P5 considered extra or irrelevant results from generative AI as harmful when under \emph{``tight time constraints''}, as they \emph{``would not want to spend time on things that are unneeded to complete the task.''} P10 wondered about balancing \emph{``how much versus how little information''} that generative AI puts into a response, and how they could control this amount of information produced to their preferences. P12 expressed appreciation for responses that were \emph{``a good balance''} of information  \emph{``between bullet points and short paragraphs''}, and \emph{``not just a two sentence answer that doesn't give any information.''}


\paragraph{Goal-satisfaction of the response.} Participants could face barriers in progressing with their task if the results only repeated what they already knew and did not add any further information, or if the results were incomplete, or incorrect, or too broad.

Some participants were suggested solutions they already knew about, but which could be useful for novices \emph{``unaware of these methods''} (P7) or \emph{``starting from scratch''} (P11). P7 requested \emph{``three more suggestions''} to elicit more unfamiliar solutions. Occasionally, the model would fail to interpret very basic and clear instructions correctly. For example, P12 was surprised that the system incorrectly applied a literary analysis framework to one story (``The Glowing Coal'') when specifically asked to apply the framework to a different story (``ATU 333, Little Red Riding Hood''). P12 wondered if the data needed to complete the task was not available to Bing Chat.




Participants also received incomplete responses from Bing Chat (P11, 13-15). P11 said they needed Bing Chat to provide justification for its choices. P13 and P15 both had replies that were useful, but incomplete since it failed to address every part of their question. E.g., one result was \emph{``not able to achieve the task''}, since it missed out the step to \emph{``convert a column''} (P13).



Other participants noted that some responses were not applicable to their specific preference, but could nonetheless be helpful in other situations (P1, 3). P14 considered a response to be \emph{``just an introduction''} to the topic and not applicable to their task. 
P4 wanted \emph{``the data to be shown in a different form.''} 
Similarly, P9 asked for a data visualization which Bing Chat provided, but P9 instead preferred a bar chart instead as it was \emph{``much more useful than a pie chart.''}

Moreover, model ``misinterpretations'' could also function as a sort of tolerance for imprecise or incorrect querying: P11 was surprised that Bing Chat ignored part of a prompt and gave what was more likely correct when P11 tried to modify a table of object detection models produced by Bing Chat by asking it to \emph{``add a column of the platforms (e.g., iOS, Android, Raspberry Pi) supported by each model''}. Bing Chat instead added a column with values like \emph{``CPU, GPU, DSP, EdgeTPU''}, which P11 realized was actually what they wanted to see in the table. P11 thought that had Bing Chat provided what was originally asked for it would have been incorrect, and instead preferred that Bing Chat intervene and recommend \emph{``corrected information''} like it had.

\paragraph{Formatting of the response.}

Another issue that participants faced was getting responses in a useful format. For example, P11 attempted to compare popular object detection models and their characteristics so they might choose the best one, but the initial reply was a bulleted list of several models and their characteristics, which made it difficult to compare between models. P11 requested Bing Chat to produce a table that specifically compared \emph{``accuracy, speed, and size''} and link to the code repository of each model. After inspecting the resultant table, P11 iterated to add columns for additional model properties. While P11 could have potentially created a detailed prompt to get a satisfactory answer with a completed table in a single step, P11 preferred to iterate and make incremental progress.

Similarly, when textual results were reformatted into a table P10 thought the results were \emph{``perfect''} since the original outputs were \emph{``very text heavy''}, but did not originally ask for a table. Thus, P10 placed the blame on Bing Chat's vague answers on the vagueness of the question they asked.


\paragraph{Intelligibility of the response.}

Finally, participants faced difficulty comprehending responses which referred to unfamiliar concepts (in a scenario where the participant was not expecting to encounter an unfamiliar concept). For example, when Bing Chat replied to a question with R functions that P5 did not know about, P5 requested an explanation of the functions and their relevance to the problem being analysed. In another example, Bing Chat recommended``Pivot Tables'' to P9, which they were unfamiliar with, but P9 said they would \emph{``just ask [Bing Chat] how to use pivot tables and for examples''} to learn more about unfamiliar concepts that generative AI recommends.

\subsubsection{Barriers to verification and trust} \label{sec:verificationbarriers} Another category of barriers was associated with the work required to assess the reliability and validity of generative AI's output, both in specific instances of AI output but also in terms of developing a mental model for the system's strengths and weaknesses in different tasks, and an overall conception of trust in the system.


\paragraph{Verification strategies} Participants developed strategies for detecting and addressing incorrect output. To understand non-working code, P1 thought they would leverage traditional resources \emph{``like textbooks''} that seemed \emph{``slightly more professional''} than Bing Chat, or ask co-workers for help. 

\begin{figure*}[t]
  \centering
  \includegraphics[width=\linewidth]{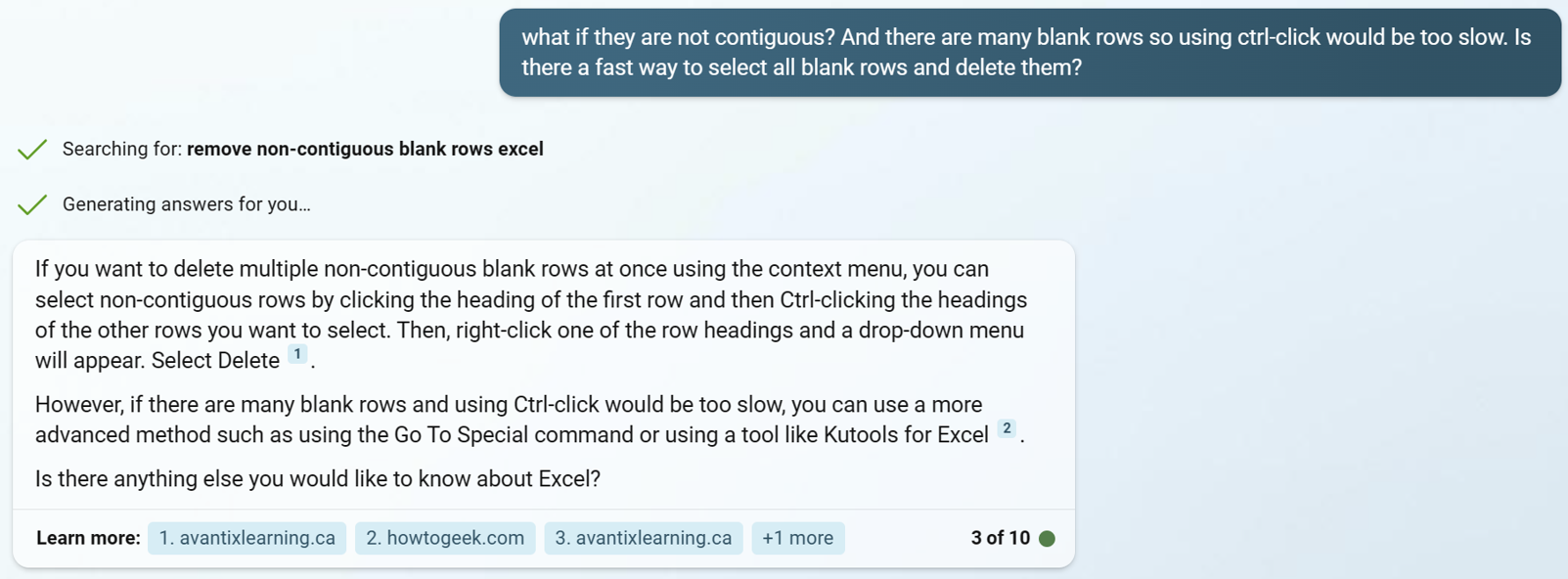}
  \caption{Bing Chat referencing UI (public design at the time of the study). The user message is in the dark blue bubble, top right. Bing Chat's response is in the white bubble, bottom left. Footnote-style superscripts indicate a supporting URL. The URLs are listed along the bottom of the chat response. A user can follow the superscripts or the links to read the web sources.}
  \Description{A screenshot of the Bing Chat referencing UI.}
  \label{fig:bing_references}
\end{figure*}

A common validation strategy was to follow the inline references (P10, 12, 14). Bing Chat provides references to the URLs from which it derives its responses using footnote-style superscripts (Figure~\ref{fig:bing_references}). During the study, P14 followed a reference link, then described a previous experience with ChatGPT where it could not present similar reference links which P14 wanted to save in EndNote. P9 also compared Bing Chat to ChatGPT, finding the citation feature \emph{``much better and more reliable''}.

Citations were seen as a fairness mechanism that \emph{``gives credit where credit is due''} (P10). However, P12 found that checking citations \emph{``becomes a process of verifying all the information it's giving you, and it might have just been quicker to find the sources yourself.''} P6 said that they will \emph{``have to verify''} each source and \emph{``use those sources to further search''}. When performing data analysis, P2 said they need to \emph{``validate that the data is from the right source''}, including the timestamps and recency of the data.

Source quality mattered. P7 preferred sites they \emph{``already trust''}, rather than unfamiliar ones.  P9 and P12 manually inspected the sources cited by Bing Chat for quality and relevance, which increased their trust of the output. P9 checked if a reference was \emph{``a scholarly article or just a website''}, preferring \emph{``trustable research''} publications, and inspected the publication date to ensure recency. P10 liked the citations, but if they were missing, they said they would just use traditional web search to verify the result themselves.



Some participants considered the seriousness of the task when deciding how much to trust and verify the response. For one task, P10 said they \emph{``trust the results, because this is such a low stakes query.''}
P12 said they would trust output if it \emph{``sounded right''} to them, unless they \emph{``really needed it to be right.''}

Verification might also involve testing and applying AI output in a different tool (P1-3, 9-11). P2 would take generated Excel formulas and \emph{``test it directly''} on their data, but P1 noted that this might be challenging without first \emph{``cleaning up the data.''} P3 also tested a generated formula \emph{``as an example''}, and then edited it to fit their needs. P4 said when they were presented with step-by-step instructions, they would \emph{``try it, and if it's not working out, do further research''} by searching online, asking colleagues for help, or watching video tutorials on YouTube. P9 had a similar approach to generated SPSS code: they first ran the code on test data to \emph{``see if it makes sense''}, before applying it to their dataset. P13 stated that \emph{``the only way to know the code is correct is to put it into an IDE.''} P3 worried about errors, which decreased their trust of generative AI, saying that they \emph{``don't rely on it [generative AI]''} and always rigorously verified any generated formulas.




\paragraph{Hallucinations} Hallucination, defined as ``generated content that is nonsensical or unfaithful to the provided source content'' \cite{ji2023hallucination}, limited generative AI's usefulness for data analysis for our participants, because it was difficult to detect (especially when the hallucination is about a domain in which the user is not an expert) and time-consuming, requiring careful and detailed attention to every part of the output. 

For example, P9 stated that they would not use it for literature reviews because of this risk. P6 felt a burden of \emph{``always having to double-check and read every line''} of the response. P11 said that while their \emph{``personal strategy is to verify everything''}, it was time-consuming and \emph{``not always possible or feasible''} to do so. Moreover, P5 described difficulty in verifying generative AI output for domains they did not \emph{``have a strong understanding in''}.


When Bing Chat started hallucinating data for P6's task, P6 said they started to \emph{``understand when you should use [generative AI] and when you shouldn't''}. P6 subsequently formed a belief that Bing Chat was not able to index copyrighted media like books, and stated that the system ought to \emph{``say `I cannot access this book or its chapters' rather than continuing to make things up''}.
P11 had an experience of \emph{``generating a function in the code that looked very authentic, but didn't exist.''} To mitigate the impact of such hallucinations, P11 aimed to \emph{``always verify all information''}, but noted that users who \emph{``blindly trust these AI tools can easily be misguided''} by hallucinations. P9 suggested that \emph{``more specific prompts to focus on a specific topic''} might address hallucinations.



\subsection{Explicit feature speculations}
\label{sec:featurespeculations}
Participants on some occasions explicitly speculated about features that would help them with sensemaking. In the traditions of HCI research deriving from sociology and cognitive psychology, study participants are not conventionally involved in the direct design of products, and as such, explicit feature requests and speculations are treated as potential evidence of a deeper underlying need which may or may not be best satisfied by implementing the feature requested. On the other hand, since we are invoking the participatory design tradition \cite{schuler1993participatory}, we are explicitly interested in participant's design speculations and consider them as first-class design contributions, at face value. We report these feature speculations in this section.

\paragraph{Application integration} Some participants saw a need for integration with the data applications they already used (P1, 2, 7, 11-13). For P13 to \emph{``be comfortable in the analysis flow of using generative AI, it would be integrated in whatever system being used on the side, and not taking up the whole screen.''} P7 thought that if they could \emph{``do it all in just Excel''}, the generative AI to have access to charts and data within the spreadsheet, reducing the effort of \emph{``going between different tabs''}. Further, P11 said that their analyses frequently ended up in slideshow presentations, so they wanted the generative AI to leverage features of one app (Excel) and place them into the final app (PowerPoint). Similarly, P6 wanted to go from chat to spreadsheet by having Bing Chat create a spreadsheet for them, avoiding a \emph{``very manual''} process of creating spreadsheets by iterating on \emph{``what categories should be included and filling out information''} (P6).

P12 believed that app-specific generative AI would \emph{``save a lot of time spent procrastinating''} such as \emph{``going down Wikipedia rabbit holes''} (e.g., exploring various related topics that are not critical to solving the task at hand). However, P1 enjoyed the broad possibilities of a general generative AI chat and worried that when leveraging generative AI within an application, the AI might limit their suggestions to operate within that application, even if a better solution might exist in another application. Instead, P1 thought that both in-app generative AI and a general generative AI would be useful, where when the in-app generative AI failed to accomplish the task, the general generative-AI could act as \emph{``the big boss who is like `alright, we'll sort this out'''}.

\paragraph{Context} We previously noted that providing context was a challenge (part of the larger set of barriers to query formulation, Section~\ref{sec:sensemaking-barriers}). Several participants offered suggestions for sharing context more easily. P1 desired a way to easily include topics and keywords of interest. P4 wanted to give negative examples, to showing what \emph{``does not exactly fit into what is wanted''} to tune the responses to be \emph{``more specific''} to the goal. P1 wished to upload their entire dataset and \emph{``have it [the AI] go''}. P5 wanted to upload \emph{``particular columns''} of their dataset as context for questions like \emph{``what can I do with this particular column''} rather than getting \emph{``generalized responses''}.


P11 described the need for chat histories they could revisit and reuse \emph{``after months''} away, to \emph{``pick up where we stopped last time and continue from there without redoing everything I did before''}. P7 wanted to go further and share chat histories with others, which might help collaborators understand the provenance of some analysis activity.


\paragraph{Formatting and modality} Participants saw a need for better intelligence and flexibility in output formatting. For example, P10 desired the data they received from Bing Chat to be in a table, which Bing Chat was able to provide after re-prompting. P10 then thought the data was \emph{``organized nicely and not overwhelming''} and could be exported easily to other applications. P11 ran into a similar issue while comparing two paragraphs, noting that placing the data in a table and comparing the columns would be \emph{``more useful''} than reading each in sequential text.

Participants also described how generative AI might go beyond text and into other modalities (P4-11, 13). Several participants saw videos, imagery, interactive maps, and other visualizations as improvements over purely textual output (P4, 10, 11, 13) depending on the problem being solved (P11). Video tutorials could provide \emph{``further clarity to see the step-by-step process''} (P4). 
P13 described example visualizations provided by Bing Chat as inspirational examples for how they could themselves visualize their data. However, P13 worried that visualizations could also be distracting and take user attention away from the text. 

On the input side, P8 also wanted to provide images and videos as part of a prompt to provide context to Bing Chat, instead of just providing text.  P11 suggested that voice interaction would feel \emph{``more natural, like talking to a human assistant.''}




\paragraph{Anthropomorphisation and social cues} 
Some participants reflected positively on Bing Chat's ability to use emojis and seem \emph{``friendly''} (P10, 13). P10 noted it was a \emph{``almost human reaction''} and said it was \emph{``nice to feel like you're talking to some sort of person or feel kind of happy [...] like texting a friend''}. However, P9 thought this style of reply \emph{``felt strange''} and was \emph{``confusing''} for them in the context of doing work with Bing Chat, since they felt like they had to make conversation with the chatbot rather than just getting answers from it.

\section{Discussion}
\subsection{Connections with related work}


\paragraph{How generative AI conversations compare to search workflows.} Participants in our study compared generative AI to traditional search workflows, finding that the linear, summarised and aggregated nature of Bing Chat responses required less effort in comparison to manually viewing multiple search results and developing a mental summary oneself (Section~\ref{sec:genai-sensemaking}). The consumer-facing positioning of the Bing Chat interface is as a complement to the more traditional Bing search engine, so to some extent this comparison is a natural one to draw, but other studies have also noted the comparison to search engines even in interfaces without such associations. For instance, studies of language model assistance in programming through code completion tools such as GitHub Copilot also find that participants cite a reduced effort in comparison to manual web search as a benefit of these tools \cite{sarkar2022programmingai}, though there are also drawbacks: due to the limited scope of sources and generation formats, language model interfaces generally offer a less media-rich experience, with fewer opportunities for learning and tangential exploration, and with fewer cues about the provenance of the results. A related observation from our participants is that search results for data analysis workflows require further work in order to adapt to the task at hand, whereas generative AI can often perform part or all of the adaptation needed. This benefit has also been observed in previous studies \cite{sarkar2022programmingai}, and it is an important benefit given that many end-user data sensemaking workflows involve such search and adaptive reuse of resources on the Web (i.e., ``transmogrification'' \cite{lau2021tweakit}).


\paragraph{Generative AI and creativity in data sensemaking.} Participants generally valued the creative potential of Bing Chat for ideation and the generation of alternative perspectives, though some participants stated a preference for first ideating and forming research questions privately (i.e., without generative AI assistance) and only using generative AI for specific data analysis tasks (Section~\ref{sec:genai-sensemaking}). At least one participant was concerned about the preservation of personal voice and style when using AI-generated text. This mix of optimism and caution has been reflected in multiple other fields, such as programming, creative writing, and visual art \cite{sarkar2023exploring}, where similarly, some aspects of creativity can be usefully attributed to the AI system, and AI can be viewed as a potential source and enhancer of creativity, but there are still important roles for humans to play, as curators, as editors, as critics, and as integrators.


\paragraph{Generative AI and common ground} A key set of challenges faced by our participants revolved around understanding and providing the context needed by Bing Chat to address their request (Section~\ref{sec:sensemaking-barriers}). Participants explicitly drew a comparison to interacting with human colleagues, where interactions were simplified to due to the vastly greater degree of shared implicit context, some deriving from the shared domain of work, others from the broader shared experience of culture and language. A concept from linguistics that captures this is the notion of \emph{common ground} \cite{stalnaker2002common}, the set of contextual presuppositions held by interlocutors that allows any speech acts to be performed and interpreted at all, without devolving into an infinite regression of ``but what does \emph{that} mean?''. Human users and generative AI do share a certain amount of common ground (deriving from the fact that generative AI behaviour is stochastic replay of real human behaviour \cite{sarkar2023enough}), but the quality of this common ground in our study was perceived as both alien and inferior to that shared between human collaborators. This aligns with the conclusions of \citet{gu2023data}, who suggest that AI assistance should be grounded in an understanding of users' current analysis plan, statistical and domain background, and overall goals; likewise, users should understand the goals of the AI assistance (e.g., to help with analysis execution, high-level planning etc.). Researchers have thus proposed to investigate how design might facilitate the notation and sharing of such contextual information without burdening the user \cite{sarkar2022explainable,gu2023data}, but to our knowledge there are no compelling solutions, and this is one of the trickier open challenges for interaction design of generative AI.


\paragraph{Folk theories and external influences} When confronted with a response that did not fit their needs or expectations, participants usually proceeded by developing a hypothesis about why the model had responded in the way it had, and adapting their next prompt accordingly, including specific strategies such as using full names of entities rather than abbreviations (e.g., ``Microsoft'' and not ``MSFT''), despite not necessarily having evidence that such hypotheses were correct, or that such strategies would be effective (Section~\ref{sec:sensemaking-barriers}). This echoes findings from other studies such as Liu, Sarkar et al. \cite{liu2023wants}, who found that participants drew from a wide range of linguistic influences, from web search to programming languages, to inform their hypotheses about how to prompt the AI system effectively. Due to the stochastic nature of generative AI, these hypotheses and consequent prompt refinements can very well produce an improved result, affirming the participants' mental model. Over time, this may result in the development of folk theories \cite{johnson1998basic} about prompting and behaviour of generative AI that may not necessarily be reliable.


\paragraph{Anthropomorphism of Generative AI in data sensemaking} Bing Chat is mildly anthropomorphised and frequently introduces emoji into its responses. Some participants noted this as a benefit as it improved the collegiality of the interaction, while others felt that it introduced an unwarranted expectation of politeness, verbosity, and conversationality on the part of the user (Section~\ref{sec:sensemaking-barriers}). This is also reflected in other studies of anthropomorphism in AI, which have found that introduction of human-like features can help users be more forgiving of a system that makes errors \cite{jensen2022human}, and improves its perceived likeability, but can be counterproductive for a system with high performance and focus on task completion \cite{de2016almost}. It is unclear from our findings whether there is a single correct approach for data sensemaking, which includes a blend of activities, some of which the system may be able to perform with high accuracy, and some not. More likely, the suitability of anthropomorphising features such as emoji appears to be dependent on the context and individual preferences.


\paragraph{Iteration and incremental progress} We noted that participants iterated with Bing Chat to incrementally build up an optimal response (Section~\ref{sec:sensemaking-barriers}), by issuing a series of prompts to slightly refine the previous response, as opposed to building up a single detailed prompt to satisfy all the requirements. This tendency to favour incremental progress has been noted in multiple previous studies of end-user interaction with AI in spreadsheets (e.g., building up a complex result through a series of intermediate columns \cite{liu2023wants}, or incrementally training a machine learning model through an ``edit, learn, guess'' loop \cite{sarkar2015interactive}). This preference for incremental interaction is similar to the motivation for direct manipulation interfaces and their property of being ``rapid, incremental, and reversible'' \cite{shneiderman1983direct}, and might be the result of the same cognitive factors that underlie the success of the direct manipulation paradigm. However, more research is needed to understand whether this is the case, and if so why, since it would appear to contradict the well-documented tendency of end-user programmers to favour the shortest path to their goal.

\paragraph{The burden of verification} Participants found that manually verifying sources was burdensome, and in some cases the work of verifying a response might be greater than the work required to conduct a web search manually (Section~\ref{sec:sensemaking-barriers}). The increased burden for users to check content has been observed in several studies (e.g., \cite{tankelevitch2023metacognitive,sarkar2023exploring,sarkar2022programmingai}). One approach to resolving this is ``co-audit'', where AI tools themselves can help to check AI-generated content \cite{gordon2023co}. What co-audit tools might look like in the context of the diverse range of data sensemaking workflows is an open research question.

\paragraph{Expertise and over-reliance} Recall that participants varied in spreadsheet usage (1 beginner, 7 experienced and basic usage, 7 experienced and advanced usage) and generative AI usage (3 never used, 1 casually use, 6 occasionally use, 5 regularly use), as well as programming experience (7 never programmed, 3 novices, 3 moderately experienced, 2 experts). In making data analysis more accessible to a wider range of non-experts through generative AI, over-reliance may become an unintended consequence (a review of the literature on AI over-reliance is given by \citet{passi2022overreliance}). We observed multiple phenomena during our study that could contribute to over-reliance, such as AI-generated output referring to concepts unfamiliar to end-users, and verification fatigue. While mitigating over-reliance was not within the scope of our study, multiple approaches have been explored such as explanations \cite{vasconcelos2023explanations}, cognitive forcing functions \cite{buccinca2021trust}, and encouraging critical thinking \cite{sarkar2024challenge,sarkar2024large}, to create appropriate reliance \cite{lee2004trust}, which is important to consider in future work. 

\paragraph{Metacognitive demands of generative AI}
Several of the issues that participants encountered align with what has been termed the `metacognitive demands' of generative AI \cite{tankelevitch2023metacognitive}. These are usability issues that reflect a need for users to have a degree of self-awareness, task decomposition, and well-adjusted confidence in their own abilities when working with generative AI systems. For example, participants struggled to formulate prompts because it was difficult to verbalise what was in their mind and break down their overall goal into sub-goals for the AI system to address---i.e., difficulties with self-awareness and task decomposition, as described and observed in other studies \cite{tankelevitch2023metacognitive, barke2023grounded,jayagopal2022exploring,dang2023choice,zamfirescu2023johnny}. Moreover, some participants found it difficult to disentangle their prompting ability from the AI system performance when certain interactions went wrong, suggesting a challenge with calibrating one's self-confidence in working with the system, as also observed in prior studies \cite{sarkar2022programmingai, zamfirescu2023johnny}. Participants' comments touched upon the role of self-confidence in verifying outputs, particularly for domains in which they have little expertise, as also observed in previous work \cite{prather2023s,weisz2021perfection}. In some cases, this was magnified by the volume of information in generated responses.


Conversely, some participants implicitly noted how the AI system provided them with metacognitive support, as outlined in \citet{tankelevitch2023metacognitive}. For example, participants commented how the system helped them think in a ``step by step'' manner, reflecting support with task decomposition. They also noted how the alternatives suggested by the system acted as inspiration when they were stuck, suggesting benefits to their metacognitive flexibility, analogously to that observed in \citet{gmeiner2023exploring}, which used human guides to support users co-creating with generative AI. 

These observations suggest that there are opportunities to design systems which explicitly provide metacognitive support to users as they approach a task, formulate prompts, and evaluate system outputs.   

\subsection{Implications for design}
\label{sec:design-implications}
Interaction design can support generative-AI assisted data sensemaking workflows (Section \ref{sec:genai-sensemaking}) by addressing barriers discovered in our study (Section \ref{sec:sensemaking-barriers}).

\emph{For query formulation:} Participants had challenges in conveying their goals and context to generative AI. These led to irrelevant, unhelpful, or partially helpful responses that required iteration to improve.
This might be addressed by a design that helps a user build more detailed prompts, e.g., proactive questions that the system provides for the user to respond to (i.e., a form of metacognitive support \cite{tankelevitch2023metacognitive}).   
Ambiguous or missing context could be detected and flagged before producing a response to avoid low quality responses.
Output formats relevant to the user's request could be recommended as prompt addenda. For example, if the user asks how to perform a specific data analysis workflow, ``step-by-step instructions'' could be suggested. This could help users improve and calibrate their confidence in their prompting ability. Designers could also explore restricted vocabularies and grammars (as opposed to unrestricted natural language queries) \cite{mu2019restricted}, or techniques such as grounded abstraction matching \cite{liu2023wants} to help users develop clearer mental models of effective querying styles.

\emph{For response inspection:} Participants also spoke about a need to verify generative AI responses for correctness, quality, and hallucinations. To do this, they inspected references provided by Bing Chat or testing code and formula suggestions. However, user expertise plays a major role in detecting incorrect output (a similar role for user expertise was observed in \citet{gu2023data}). Further, verification was effortful and time-consuming. 
Therefore, users need verification assistance, e.g., through co-auditing features \cite{gordon2023co}. The system might share strategies with the user for identifying high quality references, or assist with specifying which types of references are suitable for supporting a particular response. 
To assist users in verifying AI-generated code or formulas, the system might generate and run tests to help detect failure cases. This would speed up iteration on coding tasks. In some cases, users may not be appropriately calibrated relative to their own expertise, potentially leading to over-reliance (e.g., as in \citet{gu2023data}). Thus, as suggested in \citet{tankelevitch2023metacognitive}, there is scope for systems to prompt users to consider their own expertise and whether additional verification assistance might be helpful. 

\emph{For goal formulation:}
Participants in our study used generative AI to help them think about their data by having Bing Chat provide potential research questions or alternative analysis strategies. However, it can go further by helping users critically think about their data-driven decisions.
For example, when a user asks for AI assistance to recommend a data analysis task, the system could accompany its recommended approach with a critique of that approach outlining its potential limitations. This might prevent overreliance on the initial recommendation. The system could identify when a user's data might not be able to answer the questions they are asking, and recommend data collection strategies that would enable them to do so. To this end, \citet{gu2023data} suggest that, alongside an `analysis execution' mode, AI assistance can enter a \emph{```think' mode for specific planning suggestions, a `reflection' mode for connecting decisions and highlighting potential missed steps, and an `exploration' mode for higher-level planning suggestions''}. A step further would be to help users realise that they may not yet have a clear problem or hypothesis in mind. For example, systems can surface self-evaluation notices that encourage users to reflect on their broader aims and help them in clarifying and scoping them into concrete goals \cite{tankelevitch2023metacognitive,gmeiner2023exploring}.  

\emph{For streamlining workflows:}
Previous research has noted the challenges of cross-application workflows, particularly when using feature-rich software termed \emph{praxisware} \cite{sarkar2023simplicity}. Participants described a desire to integrate generative AI within the feature-rich applications they already use, rather than a separate experience which requires context switching between generative AI and application. This integration could help provide much of the context that our participants had trouble elaborating, as the application state already contains much of the context relevant to the task. It may also address issues with responses containing unfamiliar concepts, features, and programming languages.

However, some participants were wary of this type of integration and saw it as potentially limiting the recommendations that generative AI could provide. For example, a question asked within R studio would produce methods and code suited for doing data analysis in R, but there might be more effective strategies in other applications (e.g., Excel) that might not be provided. This limitation could be circumvented if application-specific AI systems were able to delegate queries to other applications when appropriate.

\subsection{Implications for AI research}
So far we have discussed design opportunities to improve the user experience of generative AI-assisted data analysis. This section discusses current technical developments that could positively impact the underlying issues, describe remaining gaps, and hypothesize why some issues might be addressed with foreseeable advances in technology. 

In the user journey, writing the first prompt is a significant step and our study shows that there are several issues that make query formulation difficult. Several approaches have been investigated, such as improving user prompts automatically \cite{Pryzant2023AutomaticPO, Cao2023BeautifulPromptTA} (including commercial solutions\footnote{e.g., \url{https://www.junia.ai/tools/prompt-generator}}), methods to better select prompt templates \cite{Argyle2022AnIA}, prompt banks\footnote{e.g., \url{https://github.com/f/awesome-chatgpt-prompts}} and prompt documentation\footnote{e.g., \url{https://platform.openai.com/docs/guides/prompt-engineering/prompt-engineering}}. 
A less explored avenue relates to tuning prompts such that the output is not only correct, but aligned with the users' goals. There are secondary goals when users pose a question such as learning or brainstorming (as identified in our study) and more research is needed on supporting users to write prompts that produce outputs aligned with personal goals. 

In our study, users also observed the importance and challenge of providing context. As Large Language Model (LLM) providers are continuously expanding model prompt windows (over 100,000 tokens in some cases), one might imagine that just by automatically ingesting more aspects of the user's work (e.g., the content of files on the user's filesystem, messages to collaborators, etc.) and passively relaying these to the model, we might be able to solve the context problem. Alas, several studies have shown that models struggle to identify the relevant portions in large prompts, and methods such as RAG (retrieval augmented generation) have been proposed \cite{Gao2023RetrievalAugmentedGF}. The problem worsens when the context is not inherently textual; for example, when the task needs structured knowledge via (complex) tables or knowledge bases. Despite much research effort, current evaluation still shows a significant performance gap \cite{Wang2024ChainofTableET}. 

Users identified that generative AI can provide useful and diverse responses: new datasets, complex logic, general knowledge, and inspirational ideas. Unfortunately reliability is an issue and hallucinations, or even worse inconsistent hallucinations (similar or same prompts sometimes resolve successfully, other times produce incorrect outputs), are a significant problem. Researchers have explored how to improve detection \cite{Wang2023AssessingTR}, and counteract hallucinations by grounding in verified sources \cite{Semnani2023WikiChatST}. 
No current approach can guarantee that the results generated by an LLM are correct, and research is moving towards building tools and agents that can support users to validate outputs \cite{gordon2023co}. This work is still at an early phase, but can draw from large bodies of related research such as verification, scientific reviews, and design critiques.
An interesting technical challenge is to develop an approach that lets us predict whether a generation is \emph{likely} to be correct. Because LLMs are typically optimised for next-token generation, this might require significant architectural changes. Nonetheless, this would open the door to better feedback integration in LLM generations.

\subsection{Expanding the Participatory Prompting method to other fields of interest}
\label{sec:expandingMethod}

Our research approach takes its name and inspiration from the participatory design tradition~\cite{schuler1993participatory}. That being said, the domain of data sensemaking to which we have applied it has very specific requirements that may not generalize to all use cases of the highly-flexible technology of generative AI. We believe that the method can be extended to other domains, and here make five suggestions for fundamental aspects of the method that researchers should consider.

\textit{Level of researcher intervention:} The nature of what participatory design will find depends on the interrelationship of the maturity of the technology being investigated and the level of domain expertise of the participants, but, crucially, mediated by the nature and level of activity and intervention of the researchers. Researcher mediation is a necessary part of participatory design because the approach is fundamentally about \textit{helping} end-users find agency in a context of uncertainty around technology design. Researchers may be able to take the role of a passive conduit when the participatory design process is needed to enable access to a technology that is otherwise out of reach of end-users. However, when the technology or its application are very new or involve high levels of uncertainty, researchers may need to be an active helper for participants to articulate, enact, and reflect on their own needs. This is particularly valuable in the context of generative AI, where researcher involvement enables richer, in-the-moment collection of participant data at the level of individual prompts, rather than post-hoc recollections obtained after task completion. In Sections~\ref{sec:PP-sessions} and \ref{sec:PP-turntaking} we describe the nature of our participatory prompting sessions, and how we tried to stay passive -- and in some cases could -- but often had to be more active. The more active researchers are, the more potential there is for introducing biases, but this needs to be balanced against getting reasonable results when participant uncertainty is high, and also balanced against the spirit of enabling end-user agency that is central to participatory design. As such, it is important to plan, document, and account for how active researchers need to be and actually were, so that the results can be calibrated against others in the future. 

\textit{End-user agency and ascribing agency to generative AI:} Related to the first point and researcher intervention, is that in participatory prompting, end-user agency must be more than an issue of `just' giving users a voice in the design process. The \textit{joint agency} of people and systems in participatory prompting needs to be carefully planned for, documented, and accounted for when the technology \textit{itself} is generative. That is, while researchers guide participants to see how generative technology opens up pathways for tasks hitherto difficult or impossible for them, researchers also need to guide participants on unpacking their agency in the process and track where participants ascribe agency to the technology (as our participants sometimes did in discussing how Bing Chat was part of the sensemaking loop in Section~\ref{sec:aiinthesensemakingloop}, and anthropomorphising Bing Chat in Section~\ref{sec:featurespeculations}). 

\textit{Ecological validity:} Ecological validity is the extent to which a study mimics a real situation and its findings can be generalized outside the research setting~\cite{kieffer_ecoval_2017}. While this is an issue in all research, it has a scale of relevance to participatory design depending on the domain of interest and nature of the technology. In participatory prompting studies, beyond researcher intervention mentioned above, two key aspects affecting ecological validity are: the use of participants' own materials as resources for the generative AI system, and, relatedly, the persistence of both resources and generative AI outputs across time and across technology surfaces (as noted by participants at the beginning of Section~\ref{sec:featurespeculations}). To get meaningful results, researchers will need to decide in advance how they will represent to participants the nature of ecological validity of the participatory prompting exercise and its use and persistence of resources. 

\textit{Users in groups:} Related to the third point, our study focused on one human using one generative AI system, such that the researcher was a facilitator of an individual participant's work. However, future participatory prompting studies will likely need to extend to participants acting in groups, and potentially even a hierarchy of groups (e.g. a team, the group the team belongs to, and the organisation that comprises the groups). This will entail decisions around whether participatory prompting will require exploration of each individual in a group having their own personal generative AI experience that they use in parallel to contribute to a wholly human group experience, or the group having one shared generative AI system that all can see and access serially or even some combination of both. While such group action is quite common in traditional participatory design studies, such group action maybe outside current capabilities of generative AI systems (especially group action across time and technology surfaces), necessitating some combination of real and speculative usage (or increased design fiction or Wizard-of-Oz engagement). It may also require one or more complex meta-prompts for the generative AI system so that it can (appear to) act on behalf of groups or even whole organisations. These prompts will need to be carefully designed not to misrepresent both what is feasible and what is desirable in such situations.

\textit{Domain of interest and expected outcomes:} Our study focused on sensemaking from data, which will naturally only account for a proportion of the possible workflows for the flexible technology of generative AI. The method can clearly be extended to paradigms outside data sensemaking, such as artistic creativity, idea synthesis, personal reflection on goals, Socratic dialogue, educational testing and explanation, therapeutic discussion, team project planning, and more. While some of these (e.g. education) have empirically factual outcomes that users and researchers alike could agree on, others will have outcomes more related to personal satisfaction (e.g. therapeutic discussion) or shared satisfaction (e.g. the results of a creative output), potentially some combination of both factual and satisfaction outcomes, and personal and shared outcomes (e.g. the output of a team project plan). When extending the method, then, participants and researchers need to be clear about how the nature of the domain of interest is related to the nature of preferred and expected outcomes. This is especially important given the generative AI issues around non-determinative outcomes and the potential for hallucinations, as participants voiced concerns about in Section~\ref{sec:verificationbarriers}. Such issues will be more relevant to some domains more than others. For example, verification of sources will be crucial in some domains (e.g. information analysis), others may have no sources to be verified (e.g. creative expression), the source for some will be the participant themselves (e.g. articulating and synthesising rough ideas into a coherent draft), and the `sources' for others will be the stochastic patterns of apparent human behaviour output by the models, to then be treated as satisfactory or not by participants (e.g. role-based prompting, such as asking a generative AI system to act as a travel agent or car mechanic when giving advice, planning etc.). 

\subsection{Limitations}
\label{sec:limitations}

There were limitations inherent to the Bing Chat interface which limited the kinds of behaviours we could explore. For example, some chat interfaces allow queries to be edited and re-submitted, but Bing Chat does not. If a participant wished to revise an earlier query, the best option was simply to submit the revised query as a new message, but the result might then be contaminated by the results from the previous version of the query due to the manner in which the context from the entire conversation is used in Bing Chat's responses. Nor was starting an entirely new conversation a good option, as participants often wished to continue and build on a successful conversation when revising a query. Moreover, there are features supported by other tools (e.g., ChatGPT supports plugins with varied functionality; Anthropic's Claude supports uploading and querying large documents) which we could not study. Thus, the choice of any particular tool will influence the scope of interactions which can be studied.


The set of prompting strategies was developed by trial and error, guided by the experience and subjective judgments made by a particular set of researchers. There will be differences between how different groups approach the process of developing prompting strategies, and thus this aspect of the participatory prompting process is not easily reproducible. Making this process more consistent is an important avenue for research.

As part of our protocol, each participant developed their own unique and personalised sensemaking task (Table~\ref{tbl:tasksv2}). The themes emerging from a single participant engaging with a particular task may not generalise to other participants engaging with that same task. However, for our study this was an acceptable trade-off for three reasons. First, having a wider variety of tasks improves our coverage and generalisability of insights for \emph{data-driven sensemaking} as a broad activity, which is more important than establishing generalisability for particular tasks. Second, having personal tasks developed by participants achieves ecological validity to a level that is very difficult to achieve using a synthetic suite of uniform tasks. Third, as previously mentioned, another aim of this study was to evaluate participatory prompting as a method, which more holistically and rigorously achieved using a diverse range of ecologically valid tasks.


As noted in Section~\ref{sec:sensemaking-barriers}, when encountering the researcher's mediation and pre-prepared prompting strategies, participants reflected on their own lack of awareness and perceived deficiencies in prompting strategies. Many participants described their own unmediated prompting strategies as ``too general'' and reported difficulty understanding ``where to start from.'' To some extent this validates the utility of the participatory prompting protocol; by mediating participant requests and reformulating them according to effective prompting strategies, the protocol bypasses many potential sources of frustration and shallow experiential dead-ends that might derail a 1-hour interactive study and compromise the ability to study meaningful tasks. On the other hand, this reduces the external validity of these experiences, since participants will not have access to expert mediation during real work. The amount of mediation is therefore a balance that needs to be carefully struck, to avoid over-influencing the participants' workflow; enough intervention to enable interesting and meaningful interaction but not so much that the interaction is completely different to the kind that the participant might have had on their own.

\section{Conclusion}
We studied how generative AI might affect the workflow of open-ended data analysis, i.e., sensemaking with data. We conducted participatory prompting sessions, in which participants worked with a researcher experienced in prompting strategies, to explore a data analysis problem of interest with the Bing Chat generative AI. Participants were asked to think aloud and reflect on the output at each turn of the conversation. The transcripts of the conversations with Bing Chat and the think-aloud data were thematically analysed.

We found that generative AI was useful in both the information foraging loop (by reducing the manual effort required to search for relevant information) and in the sensemaking loop (by helping ideate hypotheses, and proposing strategies to test them). On the other hand, participants faced barriers to query formulation (such as expressing their intent in detail, and determining what context needed to be shared with the system); in the utility of the responses (such as being overwhelmed by the amount of information, the response failing to meet their needs, or being unable to understand unfamiliar concepts in the response); and to verification and trust (such as the manual effort of looking for supporting information, and detailed checking for hallucinations).

The findings have design implications regarding balancing generative AI as a standalone application versus integration with other applications, helping users understand and provide context, managing the format and modality of responses, and metacognitive support. Besides viewing these as interaction design opportunities, we also highlight opportunities for technical research in machine learning to address some of these challenges. Further, we find that our data complements and extends our understanding of phenomena observed in previous research, such as the relationship of generative AI to search, creativity, common ground, folk theories, and metacognition. Finally, we reflect on the participatory prompting method as a research technique for eliciting opportunities and challenges for generative AI in knowledge workflows, consider its limitations, and how it might be applied to other domains.

\begin{acks}
We thank our participants for their time, and our reviewers for their helpful feedback.
\end{acks}

\bibliographystyle{ACM-Reference-Format}
\bibliography{references}


\end{document}

%% file: tables/tbl_participants.tex
\begin{table*}[t]
\caption{
\textbf{Participant profession} (self-reported description). 
\textbf{Spreadsheet experience} (\textbf{(1)} Little or no experience; \textbf{(2)} Some experience, but I'm still a beginner; \textbf{(3)} A lot of experience, but my use is basic; \textbf{(4)} A lot of experience, and I use some advanced features; \textbf{(5)} A lot of experience with many advanced features). 
\textbf{Programming experience} (\textbf{(1)} I have never programmed; \textbf{(2)} I have learnt a little bit but never used it; \textbf{(3)} I know enough to use it for small infrequent tasks; \textbf{(4)} I am moderately experienced and write programs regularly; \textbf{(5)} I am highly experienced). 
\textbf{Generative AI experience} (\textbf{(1)} Never heard of them; \textbf{(2)} Heard of them but haven't tried any; \textbf{(3)} Casually tried one or more; \textbf{(4)} Occasionally use one or more; \textbf{(5)} Regularly use one or more).
}
\label{tbl:demographics}
\begin{center}
\begin{tabular}{p{0.05\linewidth} p{0.30\linewidth} p{0.15\linewidth}
 p{0.15\linewidth} p{0.15\linewidth}}
P No. & 
Profession &
Spreadsheet\newline Experience & 
Programming\newline Experience &
Generative AI\newline Experience \\
\toprule

1 & PhD candidate (Anthropology) 
& 3
& 2
& 2 \\

2 & Consultant 
& 3
& 2
& 4 \\

3 & Statistician 
& 4
& 2
& 4 \\

4 & Account Officer 
& 3
& 2
& 2 \\

5 & Student (Geophysics) 
& 4 
& 4
& 5 \\

6 & Software Engineer 
& 2 
& 4
& 4 \\

7 & Data Analyst 
& 4 
& 2
& 5 \\

8 & Software Engineer 
& 4
& 4
& 5 \\

9 & Student (Informatics) 
& 3 
& 2
& 5 \\

10 & PhD student (History) 
& 3 
& 2
& 2 \\

11 & PhD candidate (Computer Science) 
& 4 
& 5
& 5 \\

12 & PhD student (Anthropology) 
& 3 
& 3
& 4 \\

13 & Computer Scientist 
& 4 
& 5
& 3 \\

14 & Student (Philosophy)
& 4 
& 3
& 4 \\

15 & Student (Nursing Science) 
& 3 
& 3
& 4 \\

\bottomrule
\end{tabular}
\end{center}
\end{table*}

%% file: tables/tbl_tasks_v2.tex
\begin{table*}
\caption{Overview of the tasks developed in collaboration with participants. Columns (left to right): participant ID, brief description of task, first prompt issued for that task to Bing Chat, and count of turns taken over the course of the task.}
\label{tbl:tasksv2}
\begin{center}
\renewcommand{\arraystretch}{1.5}
\footnotesize
\begin{tabular}{p{0.01\linewidth} p{0.18\linewidth} p{0.66\linewidth} p{0.05\linewidth}}
{ \textbf{P}} & 
{ \textbf{Description of task}} &
{ \textbf{First prompt}} & 
{ \textbf{Turns}} \\
\midrule
{ 1} 
& { Performing a literary analysis with spreadsheets}
& { I have a spreadsheet with data. In rows are the data for ``tales'' and in columns are the data for ``cooperative behaviour''. Each cell contains an example of a cooperative behaviour in a certain tale, e.g., ``brother saved brother''. I need a way to code each cell according to different categories. Explain how to use a spreadsheet for this with an example.}
& { 3} \\ \hline
{ 2}
& { Categorizing age data in a spreadsheet}
& { I have data about people with a column for their age. I need to regroup the people with age between 18-35 into categories Gen X, Gen Y, Gen Z. Explain how to use a spreadsheet for this with an example.} 
& { 2} \\ \hline

{ 3} 
& { Analysis on how discrimination impacts workplace performance} 
& { I am trying to determine the extent to discrimination affects employee performance in my company. What data is required? What online data sources may be relevant? Explain how to solve this in Excel with an example.}
& { 4} \\ \hline
{ 4} 
& { Creating a reusable expense tracker in spreadsheet}
& { I am making a form in Excel where people have to categorize their business expenses. For each expense the user has to choose a category. Instead of typing out the category I want them to be able to filter on the cell. Explain how to do this in Excel with an example.}   
& { 4} \\ \hline
{ 5} 
& { Data analysis exploration of ridesharing data} 
& { I have a dataset of rides taken on a bike sharing service. For each ride we know the rider type (casual or annual member), start and end points time and location, and bike type (classical or electric). I am trying to understand what differentiates the usage of the casual and annual members. Suggest a data analysis strategy for solving this problem using Excel and R.}
& { 4} \\ \hline
{ 6} 
& { Apartment hunting organization} 
& { I am looking for an apartment. I have a spreadsheet with the data about various apartments and the following column headers: Name, Google Rating, Location, Distance from Msft, 2Br Price, \# of Baths, Sq. ft., Move-in Date, Notes. Explain a few ways I can analyse this data in a spreadsheet to help me make my decision.}  
& { 6} \\ \hline
{ 7} 
& { Strategies to share data analyses}
& { I am a data analyst who does his analysis in Microsoft Excel. It is challenging to share the findings from my analysis because Excel files stored on my computer are not live, in the sense that I am the only one who can have access to it at any one point in time. Suggest a few solutions to this problem.} 
& { 3} \\ \hline
{ 8} 
& { Writing Javascript code}
& { Write me javascript code that adds 2 days on top of current today and has these conditions: - When it's Monday/Tuesday/Wednesday/Thursday, then the day would be Wednesday/Thursday/Friday/Saturday - When it's Friday, then the day would be Monday - When it's Saturday/Sunday, the day would be Tuesday}  
& { 6} \\ \hline
{ 9} 
& { Requesting analysis strategies for an HRI survey}
& { I am a researcher studying Human-Robot Interaction. I have data from a survey in which 100s of respondents were asked to rate their perception of two robot voices on a Likert scale from -3 to +3. I am interested in the differences between voices. Suggest a few analytical strategies for solving this problem. Suggest how to implement the strategies in Excel.}
& { 5} \\ \hline 
{ 10} 
& { Potential career paths based on background} 
& { I am a history PhD candidate with strong research, writing, and education skills. I would like to know what potential career paths I could follow based on my skills when I graduate.}
& { 7} \\ \hline
{ 11} 
& { Comparing object detection models}
& { I'm interested in the performance of object detection models. Recommend ways to compare these models. Focus on accuracy, speed in ms, and size of the model.}   
& { 3} \\ \hline
{ 12} 
& { Finding strategies for literary analysis}
& { I am interested in narrative structure. Recommend appropriate frameworks for analyzing narrative text within short stories. Cite your sources.}
& { 5} \\ \hline
{ 13} 
& { Performing data analysis on student grading data} 
& { I want to disaggregate categorical and numerical grading data by gender as part of my research study. I want to calculate the difference between the median grade and each gender. How do I do this in a spreadsheet? After this is done, I would like to export the data to latex and create a visualization of the data, can you give me the steps to perform these tasks?}  
& { 5} \\ \hline
{ 14} 
& { Discovering data about lifestyle impact on diabetic patients} 
& { Are type-2 diabetic patients compliant with lifestyle modifications? Cite your sources please.}
& { 3} \\ \hline
{ 15} 
& { Understanding what factors impact working for corporations}
& { Does university curriculum prepare people for work in corporate organizations after graduation? I want to know the answer and all the potential factors that contribute to this. Please cite your references.}
& { 7} \\
\bottomrule
\end{tabular}
\normalsize
\end{center}
\end{table*}
\renewcommand{\arraystretch}{1}

%% file: main.bbl

\begin{thebibliography}{84}


\ifx \showCODEN    \undefined \def \showCODEN     #1{\unskip}     \fi
\ifx \showDOI      \undefined \def \showDOI       #1{#1}\fi
\ifx \showISBNx    \undefined \def \showISBNx     #1{\unskip}     \fi
\ifx \showISBNxiii \undefined \def \showISBNxiii  #1{\unskip}     \fi
\ifx \showISSN     \undefined \def \showISSN      #1{\unskip}     \fi
\ifx \showLCCN     \undefined \def \showLCCN      #1{\unskip}     \fi
\ifx \shownote     \undefined \def \shownote      #1{#1}          \fi
\ifx \showarticletitle \undefined \def \showarticletitle #1{#1}   \fi
\ifx \showURL      \undefined \def \showURL       {\relax}        \fi
\providecommand\bibfield[2]{#2}
\providecommand\bibinfo[2]{#2}
\providecommand\natexlab[1]{#1}
\providecommand\showeprint[2][]{arXiv:#2}

\bibitem[Argyle et~al\mbox{.}(2022)]%
        {Argyle2022AnIA}
\bibfield{author}{\bibinfo{person}{Lisa~P. Argyle}, \bibinfo{person}{E. Busby}, \bibinfo{person}{Nancy Fulda}, \bibinfo{person}{Joshua~R Gubler}, \bibinfo{person}{Christopher Rytting}, \bibinfo{person}{Taylor Sorensen}, {and} \bibinfo{person}{David Wingate}.} \bibinfo{year}{2022}\natexlab{}.
\newblock \showarticletitle{An Information-theoretic Approach to Prompt Engineering Without Ground Truth Labels}.
\newblock \bibinfo{journal}{\emph{Political Analysis}}  \bibinfo{volume}{31} (\bibinfo{year}{2022}), \bibinfo{pages}{337 -- 351}.
\newblock
\urldef\tempurl%
\url{https://api.semanticscholar.org/CorpusID:252280474}
\showURL{%
\tempurl}


\bibitem[Barke et~al\mbox{.}(2023)]%
        {barke2023grounded}
\bibfield{author}{\bibinfo{person}{Shraddha Barke}, \bibinfo{person}{Michael~B James}, {and} \bibinfo{person}{Nadia Polikarpova}.} \bibinfo{year}{2023}\natexlab{}.
\newblock \showarticletitle{Grounded copilot: How programmers interact with code-generating models}.
\newblock \bibinfo{journal}{\emph{Proceedings of the ACM on Programming Languages}} \bibinfo{volume}{7}, \bibinfo{number}{OOPSLA1} (\bibinfo{year}{2023}), \bibinfo{pages}{85--111}.
\newblock


\bibitem[Beyer and Holtzblatt(1997)]%
        {Beyer1997}
\bibfield{author}{\bibinfo{person}{Hugh Beyer} {and} \bibinfo{person}{Karen Holtzblatt}.} \bibinfo{year}{1997}\natexlab{}.
\newblock \bibinfo{booktitle}{\emph{Contextual Design: Defining Customer-Centered Systems}}.
\newblock \bibinfo{publisher}{Morgan Kaufmann Publishers Inc.}, \bibinfo{address}{San Francisco, CA, USA}.
\newblock
\showISBNx{9780080503042}


\bibitem[Braun and Clarke(2006)]%
        {braun2006using}
\bibfield{author}{\bibinfo{person}{Virginia Braun} {and} \bibinfo{person}{Victoria Clarke}.} \bibinfo{year}{2006}\natexlab{}.
\newblock \showarticletitle{Using thematic analysis in psychology}.
\newblock \bibinfo{journal}{\emph{Qualitative research in psychology}} \bibinfo{volume}{3}, \bibinfo{number}{2} (\bibinfo{year}{2006}), \bibinfo{pages}{77--101}.
\newblock


\bibitem[Bu{\c{c}}inca et~al\mbox{.}(2021)]%
        {buccinca2021trust}
\bibfield{author}{\bibinfo{person}{Zana Bu{\c{c}}inca}, \bibinfo{person}{Maja~Barbara Malaya}, {and} \bibinfo{person}{Krzysztof~Z Gajos}.} \bibinfo{year}{2021}\natexlab{}.
\newblock \showarticletitle{To trust or to think: cognitive forcing functions can reduce overreliance on AI in AI-assisted decision-making}.
\newblock \bibinfo{journal}{\emph{Proceedings of the ACM on Human-Computer Interaction}} \bibinfo{volume}{5}, \bibinfo{number}{CSCW1} (\bibinfo{year}{2021}), \bibinfo{pages}{1--21}.
\newblock


\bibitem[Cao et~al\mbox{.}(2023)]%
        {Cao2023BeautifulPromptTA}
\bibfield{author}{\bibinfo{person}{Tingfeng Cao}, \bibinfo{person}{Chengyu Wang}, \bibinfo{person}{Bingyan Liu}, \bibinfo{person}{Ziheng Wu}, \bibinfo{person}{Jinhui Zhu}, {and} \bibinfo{person}{Jun Huang}.} \bibinfo{year}{2023}\natexlab{}.
\newblock \showarticletitle{BeautifulPrompt: Towards Automatic Prompt Engineering for Text-to-Image Synthesis}. In \bibinfo{booktitle}{\emph{Conference on Empirical Methods in Natural Language Processing}}.
\newblock
\urldef\tempurl%
\url{https://api.semanticscholar.org/CorpusID:265150243}
\showURL{%
\tempurl}


\bibitem[Chalhoub and Sarkar(2022)]%
        {chalhoub2022freedom}
\bibfield{author}{\bibinfo{person}{George Chalhoub} {and} \bibinfo{person}{Advait Sarkar}.} \bibinfo{year}{2022}\natexlab{}.
\newblock \showarticletitle{“It’s Freedom to Put Things Where My Mind Wants”: Understanding and Improving the User Experience of Structuring Data in Spreadsheets}. In \bibinfo{booktitle}{\emph{Proceedings of the 2022 CHI Conference on Human Factors in Computing Systems}} (New Orleans, LA, USA) \emph{(\bibinfo{series}{CHI '22})}. \bibinfo{publisher}{Association for Computing Machinery}, \bibinfo{address}{New York, NY, USA}, Article \bibinfo{articleno}{585}, \bibinfo{numpages}{24}~pages.
\newblock
\showISBNx{9781450391573}
\urldef\tempurl%
\url{https://doi.org/10.1145/3491102.3501833}
\showDOI{\tempurl}


\bibitem[Chattopadhyay et~al\mbox{.}(2023)]%
        {chattopadhyay2023make}
\bibfield{author}{\bibinfo{person}{Souti Chattopadhyay}, \bibinfo{person}{Zixuan Feng}, \bibinfo{person}{Emily Arteaga}, \bibinfo{person}{Audrey Au}, \bibinfo{person}{Gonzalo Ramos}, \bibinfo{person}{Titus Barik}, {and} \bibinfo{person}{Anita Sarma}.} \bibinfo{year}{2023}\natexlab{}.
\newblock \showarticletitle{Make It Make Sense! Understanding and Facilitating Sensemaking in Computational Notebooks}.
\newblock \bibinfo{journal}{\emph{arXiv preprint arXiv:2312.11431}} (\bibinfo{year}{2023}).
\newblock


\bibitem[Chen et~al\mbox{.}(2023)]%
        {chen2023whatsnext}
\bibfield{author}{\bibinfo{person}{Chen Chen}, \bibinfo{person}{Jane Hoffswell}, \bibinfo{person}{Shunan Guo}, \bibinfo{person}{Ryan Rossi}, \bibinfo{person}{Yeuk-Yin Chan}, \bibinfo{person}{Fan Du}, \bibinfo{person}{Eunyee Koh}, {and} \bibinfo{person}{Zhicheng Liu}.} \bibinfo{year}{2023}\natexlab{}.
\newblock \showarticletitle{WHATSNEXT: Guidance-enriched Exploratory Data Analysis with Interactive, Low-Code Notebooks}. In \bibinfo{booktitle}{\emph{2023 IEEE Symposium on Visual Languages and Human-Centric Computing (VL/HCC)}}. IEEE, \bibinfo{pages}{209--214}.
\newblock


\bibitem[Dang et~al\mbox{.}(2023)]%
        {dang2023choice}
\bibfield{author}{\bibinfo{person}{Hai Dang}, \bibinfo{person}{Sven Goller}, \bibinfo{person}{Florian Lehmann}, {and} \bibinfo{person}{Daniel Buschek}.} \bibinfo{year}{2023}\natexlab{}.
\newblock \showarticletitle{Choice over control: How users write with large language models using diegetic and non-diegetic prompting}. In \bibinfo{booktitle}{\emph{Proceedings of the 2023 CHI Conference on Human Factors in Computing Systems}}. \bibinfo{pages}{1--17}.
\newblock


\bibitem[De~Visser et~al\mbox{.}(2016)]%
        {de2016almost}
\bibfield{author}{\bibinfo{person}{Ewart~J De~Visser}, \bibinfo{person}{Samuel~S Monfort}, \bibinfo{person}{Ryan McKendrick}, \bibinfo{person}{Melissa~AB Smith}, \bibinfo{person}{Patrick~E McKnight}, \bibinfo{person}{Frank Krueger}, {and} \bibinfo{person}{Raja Parasuraman}.} \bibinfo{year}{2016}\natexlab{}.
\newblock \showarticletitle{Almost human: Anthropomorphism increases trust resilience in cognitive agents.}
\newblock \bibinfo{journal}{\emph{Journal of Experimental Psychology: Applied}} \bibinfo{volume}{22}, \bibinfo{number}{3} (\bibinfo{year}{2016}), \bibinfo{pages}{331}.
\newblock


\bibitem[Dorton and Hall(2021)]%
        {dorton2021collaborative}
\bibfield{author}{\bibinfo{person}{Stephen~L Dorton} {and} \bibinfo{person}{Robert~A Hall}.} \bibinfo{year}{2021}\natexlab{}.
\newblock \showarticletitle{Collaborative human-AI sensemaking for intelligence analysis}. In \bibinfo{booktitle}{\emph{International conference on human-computer interaction}}. Springer, \bibinfo{pages}{185--201}.
\newblock


\bibitem[Engels and Erwig(2005)]%
        {engels2005classsheets}
\bibfield{author}{\bibinfo{person}{Gregor Engels} {and} \bibinfo{person}{Martin Erwig}.} \bibinfo{year}{2005}\natexlab{}.
\newblock \showarticletitle{ClassSheets: automatic generation of spreadsheet applications from object-oriented specifications}. In \bibinfo{booktitle}{\emph{Proceedings of the 20th IEEE/ACM international Conference on Automated software engineering}}. \bibinfo{pages}{124--133}.
\newblock


\bibitem[Ferdowsi et~al\mbox{.}(2023)]%
        {ferdowsi2023coldeco}
\bibfield{author}{\bibinfo{person}{Kasra Ferdowsi}, \bibinfo{person}{Jack Williams}, \bibinfo{person}{Ian Drosos}, \bibinfo{person}{Andrew~D. Gordon}, \bibinfo{person}{Carina Negreanu}, \bibinfo{person}{Nadia Polikarpova}, \bibinfo{person}{Advait Sarkar}, {and} \bibinfo{person}{Benjamin Zorn}.} \bibinfo{year}{2023}\natexlab{}.
\newblock \showarticletitle{COLDECO: An End User Spreadsheet Inspection Tool for AI-Generated Code}. In \bibinfo{booktitle}{\emph{2023 IEEE Symposium on Visual Languages and Human-Centric Computing (VL/HCC)}}. \bibinfo{pages}{82--91}.
\newblock
\urldef\tempurl%
\url{https://doi.org/10.1109/VL-HCC57772.2023.00017}
\showDOI{\tempurl}


\bibitem[Gao et~al\mbox{.}(2023)]%
        {Gao2023RetrievalAugmentedGF}
\bibfield{author}{\bibinfo{person}{Yunfan Gao}, \bibinfo{person}{Yun Xiong}, \bibinfo{person}{Xinyu Gao}, \bibinfo{person}{Kangxiang Jia}, \bibinfo{person}{Jinliu Pan}, \bibinfo{person}{Yuxi Bi}, \bibinfo{person}{Yi Dai}, \bibinfo{person}{Jiawei Sun}, \bibinfo{person}{Qianyu Guo}, \bibinfo{person}{Meng Wang}, {and} \bibinfo{person}{Haofen Wang}.} \bibinfo{year}{2023}\natexlab{}.
\newblock \showarticletitle{Retrieval-Augmented Generation for Large Language Models: A Survey}.
\newblock \bibinfo{journal}{\emph{ArXiv}}  \bibinfo{volume}{abs/2312.10997} (\bibinfo{year}{2023}).
\newblock
\urldef\tempurl%
\url{https://api.semanticscholar.org/CorpusID:266359151}
\showURL{%
\tempurl}


\bibitem[Gmeiner et~al\mbox{.}(2023)]%
        {gmeiner2023exploring}
\bibfield{author}{\bibinfo{person}{Frederic Gmeiner}, \bibinfo{person}{Humphrey Yang}, \bibinfo{person}{Lining Yao}, \bibinfo{person}{Kenneth Holstein}, {and} \bibinfo{person}{Nikolas Martelaro}.} \bibinfo{year}{2023}\natexlab{}.
\newblock \showarticletitle{Exploring Challenges and Opportunities to Support Designers in Learning to Co-create with AI-based Manufacturing Design Tools}. In \bibinfo{booktitle}{\emph{Proceedings of the 2023 CHI Conference on Human Factors in Computing Systems}}. \bibinfo{pages}{1--20}.
\newblock


\bibitem[Gordon et~al\mbox{.}(2023)]%
        {gordon2023co}
\bibfield{author}{\bibinfo{person}{Andrew~D Gordon}, \bibinfo{person}{Carina Negreanu}, \bibinfo{person}{Jos{\'e} Cambronero}, \bibinfo{person}{Rasika Chakravarthy}, \bibinfo{person}{Ian Drosos}, \bibinfo{person}{Hao Fang}, \bibinfo{person}{Bhaskar Mitra}, \bibinfo{person}{Hannah Richardson}, \bibinfo{person}{Advait Sarkar}, \bibinfo{person}{Stephanie Simmons}, {et~al\mbox{.}}} \bibinfo{year}{2023}\natexlab{}.
\newblock \showarticletitle{Co-audit: tools to help humans double-check AI-generated content}.
\newblock \bibinfo{journal}{\emph{arXiv preprint arXiv:2310.01297}} (\bibinfo{year}{2023}).
\newblock


\bibitem[Grigoreanu et~al\mbox{.}(2012)]%
        {grigoreanu2012end}
\bibfield{author}{\bibinfo{person}{Valentina Grigoreanu}, \bibinfo{person}{Margaret Burnett}, \bibinfo{person}{Susan Wiedenbeck}, \bibinfo{person}{Jill Cao}, \bibinfo{person}{Kyle Rector}, {and} \bibinfo{person}{Irwin Kwan}.} \bibinfo{year}{2012}\natexlab{}.
\newblock \showarticletitle{End-user debugging strategies: A sensemaking perspective}.
\newblock \bibinfo{journal}{\emph{ACM Transactions on Computer-Human Interaction (TOCHI)}} \bibinfo{volume}{19}, \bibinfo{number}{1} (\bibinfo{year}{2012}), \bibinfo{pages}{1--28}.
\newblock


\bibitem[Gu et~al\mbox{.}(2023a)]%
        {gu2023data}
\bibfield{author}{\bibinfo{person}{Ken Gu}, \bibinfo{person}{Madeleine Grunde-McLaughlin}, \bibinfo{person}{Andrew~M McNutt}, \bibinfo{person}{Jeffrey Heer}, {and} \bibinfo{person}{Tim Althoff}.} \bibinfo{year}{2023}\natexlab{a}.
\newblock \showarticletitle{How Do Data Analysts Respond to AI Assistance? A Wizard-of-Oz Study}.
\newblock \bibinfo{journal}{\emph{arXiv preprint arXiv:2309.10108}} (\bibinfo{year}{2023}).
\newblock


\bibitem[Gu et~al\mbox{.}(2023b)]%
        {gu2023analysts}
\bibfield{author}{\bibinfo{person}{Ken Gu}, \bibinfo{person}{Ruoxi Shang}, \bibinfo{person}{Tim Althoff}, \bibinfo{person}{Chenglong Wang}, {and} \bibinfo{person}{Steven~M. Drucker}.} \bibinfo{year}{2023}\natexlab{b}.
\newblock \bibinfo{title}{How Do Analysts Understand and Verify AI-Assisted Data Analyses?}
\newblock
\newblock
\showeprint[arxiv]{2309.10947}~[cs.HC]


\bibitem[Gulwani(2011)]%
        {gulwani2011automating}
\bibfield{author}{\bibinfo{person}{Sumit Gulwani}.} \bibinfo{year}{2011}\natexlab{}.
\newblock \showarticletitle{Automating string processing in spreadsheets using input-output examples}.
\newblock \bibinfo{journal}{\emph{ACM Sigplan Notices}} \bibinfo{volume}{46}, \bibinfo{number}{1} (\bibinfo{year}{2011}), \bibinfo{pages}{317--330}.
\newblock


\bibitem[Horvath et~al\mbox{.}(2022)]%
        {horvath2022using}
\bibfield{author}{\bibinfo{person}{Amber Horvath}, \bibinfo{person}{Brad Myers}, \bibinfo{person}{Andrew Macvean}, {and} \bibinfo{person}{Imtiaz Rahman}.} \bibinfo{year}{2022}\natexlab{}.
\newblock \showarticletitle{Using Annotations for Sensemaking About Code}. In \bibinfo{booktitle}{\emph{Proceedings of the 35th Annual ACM Symposium on User Interface Software and Technology}}. \bibinfo{pages}{1--16}.
\newblock


\bibitem[Jayagopal et~al\mbox{.}(2022)]%
        {jayagopal2022exploring}
\bibfield{author}{\bibinfo{person}{Dhanya Jayagopal}, \bibinfo{person}{Justin Lubin}, {and} \bibinfo{person}{Sarah~E Chasins}.} \bibinfo{year}{2022}\natexlab{}.
\newblock \showarticletitle{Exploring the learnability of program synthesizers by novice programmers}. In \bibinfo{booktitle}{\emph{Proceedings of the 35th Annual ACM Symposium on User Interface Software and Technology}}. \bibinfo{pages}{1--15}.
\newblock


\bibitem[Jensen and Khan(2022)]%
        {jensen2022human}
\bibfield{author}{\bibinfo{person}{Theodore Jensen} {and} \bibinfo{person}{Mohammad Maifi~Hasan Khan}.} \bibinfo{year}{2022}\natexlab{}.
\newblock \showarticletitle{I’m Only Human: The Effects of Trust Dampening by Anthropomorphic Agents}. In \bibinfo{booktitle}{\emph{International Conference on Human-Computer Interaction}}. Springer, \bibinfo{pages}{285--306}.
\newblock


\bibitem[Ji et~al\mbox{.}(2023)]%
        {ji2023hallucination}
\bibfield{author}{\bibinfo{person}{Ziwei Ji}, \bibinfo{person}{Nayeon Lee}, \bibinfo{person}{Rita Frieske}, \bibinfo{person}{Tiezheng Yu}, \bibinfo{person}{Dan Su}, \bibinfo{person}{Yan Xu}, \bibinfo{person}{Etsuko Ishii}, \bibinfo{person}{Ye~Jin Bang}, \bibinfo{person}{Andrea Madotto}, {and} \bibinfo{person}{Pascale Fung}.} \bibinfo{year}{2023}\natexlab{}.
\newblock \showarticletitle{Survey of Hallucination in Natural Language Generation}.
\newblock \bibinfo{journal}{\emph{ACM Comput. Surv.}} \bibinfo{volume}{55}, \bibinfo{number}{12}, Article \bibinfo{articleno}{248} (\bibinfo{date}{mar} \bibinfo{year}{2023}), \bibinfo{numpages}{38}~pages.
\newblock
\showISSN{0360-0300}
\urldef\tempurl%
\url{https://doi.org/10.1145/3571730}
\showDOI{\tempurl}


\bibitem[Joharizadeh et~al\mbox{.}(2020)]%
        {joharizadeh2020gridlets}
\bibfield{author}{\bibinfo{person}{Nima Joharizadeh}, \bibinfo{person}{Advait Sarkar}, \bibinfo{person}{Andrew~D. Gordon}, {and} \bibinfo{person}{Jack Williams}.} \bibinfo{year}{2020}\natexlab{}.
\newblock \showarticletitle{Gridlets: Reusing Spreadsheet Grids}. In \bibinfo{booktitle}{\emph{Extended Abstracts of the 2020 CHI Conference on Human Factors in Computing Systems}} (Honolulu, HI, USA) \emph{(\bibinfo{series}{CHI EA '20})}. \bibinfo{publisher}{Association for Computing Machinery}, \bibinfo{address}{New York, NY, USA}, \bibinfo{pages}{1–7}.
\newblock
\showISBNx{9781450368193}
\urldef\tempurl%
\url{https://doi.org/10.1145/3334480.3382806}
\showDOI{\tempurl}


\bibitem[Johnson-Laird and Oatley(1998)]%
        {johnson1998basic}
\bibfield{author}{\bibinfo{person}{Philip~N Johnson-Laird} {and} \bibinfo{person}{Keith Oatley}.} \bibinfo{year}{1998}\natexlab{}.
\newblock \showarticletitle{Basic emotions, rationality, and folk theory}.
\newblock In \bibinfo{booktitle}{\emph{Consciousness and Emotion in Cognitive Science}}. \bibinfo{publisher}{Routledge}, \bibinfo{pages}{289--311}.
\newblock


\bibitem[Jones et~al\mbox{.}(2003)]%
        {jones2003user}
\bibfield{author}{\bibinfo{person}{Simon~Peyton Jones}, \bibinfo{person}{Alan Blackwell}, {and} \bibinfo{person}{Margaret Burnett}.} \bibinfo{year}{2003}\natexlab{}.
\newblock \showarticletitle{A user-centred approach to functions in Excel}. In \bibinfo{booktitle}{\emph{Proceedings of the eighth ACM SIGPLAN international conference on Functional programming}}. \bibinfo{pages}{165--176}.
\newblock


\bibitem[{Karl E. Weick}(1969)]%
        {weick_socialpsychorgz_1969}
\bibfield{author}{\bibinfo{person}{{Karl E. Weick}}.} \bibinfo{year}{1969}\natexlab{}.
\newblock \bibinfo{booktitle}{\emph{The {Social} {Psychology} of {Organizing}}}.
\newblock \bibinfo{publisher}{Addison Wesley}, \bibinfo{address}{Reading, MA}.
\newblock


\bibitem[{Karl E. Weick}(1995)]%
        {weick_sensemaking_1995}
\bibfield{author}{\bibinfo{person}{{Karl E. Weick}}.} \bibinfo{year}{1995}\natexlab{}.
\newblock \bibinfo{booktitle}{\emph{Sensemaking in {Organizations}}}.
\newblock \bibinfo{publisher}{SAGE Publications}, \bibinfo{address}{Thousand Oaks, CA}.
\newblock


\bibitem[Kieffer(2017)]%
        {kieffer_ecoval_2017}
\bibfield{author}{\bibinfo{person}{Suzanne Kieffer}.} \bibinfo{year}{2017}\natexlab{}.
\newblock \showarticletitle{{ECOVAL}: {Ecological} {Validity} of {Cues} and {Representative} {Design} in {User} {Experience} {Evaluations}}.
\newblock \bibinfo{journal}{\emph{AIS Transactions on Human-Computer Interaction}} \bibinfo{volume}{9}, \bibinfo{number}{2} (\bibinfo{date}{June} \bibinfo{year}{2017}), \bibinfo{pages}{149--172}.
\newblock
\showISSN{1944-3900}
\urldef\tempurl%
\url{https://aisel.aisnet.org/thci/vol9/iss2/4}
\showURL{%
\tempurl}


\bibitem[Ko et~al\mbox{.}(2011)]%
        {ko2011state}
\bibfield{author}{\bibinfo{person}{Amy~J Ko}, \bibinfo{person}{Robin Abraham}, \bibinfo{person}{Laura Beckwith}, \bibinfo{person}{Alan Blackwell}, \bibinfo{person}{Margaret Burnett}, \bibinfo{person}{Martin Erwig}, \bibinfo{person}{Chris Scaffidi}, \bibinfo{person}{Joseph Lawrance}, \bibinfo{person}{Henry Lieberman}, \bibinfo{person}{Brad Myers}, {et~al\mbox{.}}} \bibinfo{year}{2011}\natexlab{}.
\newblock \showarticletitle{The state of the art in end-user software engineering}.
\newblock \bibinfo{journal}{\emph{ACM Computing Surveys (CSUR)}} \bibinfo{volume}{43}, \bibinfo{number}{3} (\bibinfo{year}{2011}), \bibinfo{pages}{1--44}.
\newblock


\bibitem[Lau et~al\mbox{.}(2021)]%
        {lau2021tweakit}
\bibfield{author}{\bibinfo{person}{Sam Lau}, \bibinfo{person}{Sruti~Srinivasa Srinivasa~Ragavan}, \bibinfo{person}{Ken Milne}, \bibinfo{person}{Titus Barik}, {and} \bibinfo{person}{Advait Sarkar}.} \bibinfo{year}{2021}\natexlab{}.
\newblock \showarticletitle{TweakIt: Supporting End-User Programmers Who Transmogrify Code}. In \bibinfo{booktitle}{\emph{Proceedings of the 2021 CHI Conference on Human Factors in Computing Systems}} (Yokohama, Japan) \emph{(\bibinfo{series}{CHI '21})}. \bibinfo{publisher}{Association for Computing Machinery}, \bibinfo{address}{New York, NY, USA}, Article \bibinfo{articleno}{311}, \bibinfo{numpages}{12}~pages.
\newblock
\showISBNx{9781450380966}
\urldef\tempurl%
\url{https://doi.org/10.1145/3411764.3445265}
\showDOI{\tempurl}


\bibitem[Lee and See(2004)]%
        {lee2004trust}
\bibfield{author}{\bibinfo{person}{John~D Lee} {and} \bibinfo{person}{Katrina~A See}.} \bibinfo{year}{2004}\natexlab{}.
\newblock \showarticletitle{Trust in automation: Designing for appropriate reliance}.
\newblock \bibinfo{journal}{\emph{Human factors}} \bibinfo{volume}{46}, \bibinfo{number}{1} (\bibinfo{year}{2004}), \bibinfo{pages}{50--80}.
\newblock


\bibitem[Lee et~al\mbox{.}(2015)]%
        {lee2015people}
\bibfield{author}{\bibinfo{person}{Sukwon Lee}, \bibinfo{person}{Sung-Hee Kim}, \bibinfo{person}{Ya-Hsin Hung}, \bibinfo{person}{Heidi Lam}, \bibinfo{person}{Youn-ah Kang}, {and} \bibinfo{person}{Ji~Soo Yi}.} \bibinfo{year}{2015}\natexlab{}.
\newblock \showarticletitle{How do people make sense of unfamiliar visualizations?: A grounded model of novice's information visualization sensemaking}.
\newblock \bibinfo{journal}{\emph{IEEE transactions on visualization and computer graphics}} \bibinfo{volume}{22}, \bibinfo{number}{1} (\bibinfo{year}{2015}), \bibinfo{pages}{499--508}.
\newblock


\bibitem[Lewis and Wharton(1997)]%
        {lewis1997cognitive}
\bibfield{author}{\bibinfo{person}{Clayton Lewis} {and} \bibinfo{person}{Cathleen Wharton}.} \bibinfo{year}{1997}\natexlab{}.
\newblock \showarticletitle{Cognitive walkthroughs}.
\newblock In \bibinfo{booktitle}{\emph{Handbook of human-computer interaction}}. \bibinfo{publisher}{Elsevier}, \bibinfo{pages}{717--732}.
\newblock


\bibitem[Li et~al\mbox{.}(2023)]%
        {li2023edassistant}
\bibfield{author}{\bibinfo{person}{Xingjun Li}, \bibinfo{person}{Yizhi Zhang}, \bibinfo{person}{Justin Leung}, \bibinfo{person}{Chengnian Sun}, {and} \bibinfo{person}{Jian Zhao}.} \bibinfo{year}{2023}\natexlab{}.
\newblock \showarticletitle{EDAssistant: Supporting Exploratory Data Analysis in Computational Notebooks with In Situ Code Search and Recommendation}.
\newblock \bibinfo{journal}{\emph{ACM Trans. Interact. Intell. Syst.}} \bibinfo{volume}{13}, \bibinfo{number}{1}, Article \bibinfo{articleno}{1} (\bibinfo{date}{mar} \bibinfo{year}{2023}), \bibinfo{numpages}{27}~pages.
\newblock
\showISSN{2160-6455}
\urldef\tempurl%
\url{https://doi.org/10.1145/3545995}
\showDOI{\tempurl}


\bibitem[Liu et~al\mbox{.}(2023)]%
        {liu2023wants}
\bibfield{author}{\bibinfo{person}{Michael~Xieyang Liu}, \bibinfo{person}{Advait Sarkar}, \bibinfo{person}{Carina Negreanu}, \bibinfo{person}{Benjamin Zorn}, \bibinfo{person}{Jack Williams}, \bibinfo{person}{Neil Toronto}, {and} \bibinfo{person}{Andrew~D. Gordon}.} \bibinfo{year}{2023}\natexlab{}.
\newblock \showarticletitle{“What It Wants Me To Say”: Bridging the Abstraction Gap Between End-User Programmers and Code-Generating Large Language Models}. In \bibinfo{booktitle}{\emph{Proceedings of the 2023 CHI Conference on Human Factors in Computing Systems}} (Hamburg, Germany) \emph{(\bibinfo{series}{CHI '23})}. \bibinfo{publisher}{Association for Computing Machinery}, \bibinfo{address}{New York, NY, USA}, Article \bibinfo{articleno}{598}, \bibinfo{numpages}{31}~pages.
\newblock
\showISBNx{9781450394215}
\urldef\tempurl%
\url{https://doi.org/10.1145/3544548.3580817}
\showDOI{\tempurl}


\bibitem[Lupton(2016)]%
        {lupton2016quantified}
\bibfield{author}{\bibinfo{person}{Deborah Lupton}.} \bibinfo{year}{2016}\natexlab{}.
\newblock \bibinfo{booktitle}{\emph{The quantified self}}.
\newblock \bibinfo{publisher}{John Wiley \& Sons}.
\newblock


\bibitem[M{\u{a}}r{\u{a}}{\c{s}}oiu et~al\mbox{.}(2016)]%
        {muaruacsoiu2016clarifying}
\bibfield{author}{\bibinfo{person}{Mariana M{\u{a}}r{\u{a}}{\c{s}}oiu}, \bibinfo{person}{Alan~F Blackwell}, \bibinfo{person}{Advait Sarkar}, {and} \bibinfo{person}{Martin Spott}.} \bibinfo{year}{2016}\natexlab{}.
\newblock \showarticletitle{Clarifying hypotheses by sketching data}. In \bibinfo{booktitle}{\emph{Proceedings of the Eurographics/IEEE VGTC Conference on Visualization: Short Papers}}. \bibinfo{pages}{125--129}.
\newblock


\bibitem[Mccutchen et~al\mbox{.}(2020)]%
        {mccutchen2020elastic}
\bibfield{author}{\bibinfo{person}{Matt Mccutchen}, \bibinfo{person}{Judith Borghouts}, \bibinfo{person}{Andrew~D Gordon}, \bibinfo{person}{Simon~Peyton Jones}, {and} \bibinfo{person}{Advait Sarkar}.} \bibinfo{year}{2020}\natexlab{}.
\newblock \showarticletitle{Elastic sheet-defined functions: Generalising spreadsheet functions to variable-size input arrays}.
\newblock \bibinfo{journal}{\emph{Journal of Functional Programming}}  \bibinfo{volume}{30} (\bibinfo{year}{2020}), \bibinfo{pages}{e26}.
\newblock


\bibitem[McDonald et~al\mbox{.}(2019)]%
        {mcdonald2019reliability}
\bibfield{author}{\bibinfo{person}{Nora McDonald}, \bibinfo{person}{Sarita Schoenebeck}, {and} \bibinfo{person}{Andrea Forte}.} \bibinfo{year}{2019}\natexlab{}.
\newblock \showarticletitle{Reliability and inter-rater reliability in qualitative research: Norms and guidelines for CSCW and HCI practice}.
\newblock \bibinfo{journal}{\emph{Proceedings of the ACM on human-computer interaction}} \bibinfo{volume}{3}, \bibinfo{number}{CSCW} (\bibinfo{year}{2019}), \bibinfo{pages}{1--23}.
\newblock


\bibitem[McNutt et~al\mbox{.}(2023)]%
        {mcnutt2023design}
\bibfield{author}{\bibinfo{person}{Andrew~M McNutt}, \bibinfo{person}{Chenglong Wang}, \bibinfo{person}{Robert~A Deline}, {and} \bibinfo{person}{Steven~M Drucker}.} \bibinfo{year}{2023}\natexlab{}.
\newblock \showarticletitle{On the design of ai-powered code assistants for notebooks}. In \bibinfo{booktitle}{\emph{Proceedings of the 2023 CHI Conference on Human Factors in Computing Systems}}. \bibinfo{pages}{1--16}.
\newblock


\bibitem[Mu and Sarkar(2019)]%
        {mu2019restricted}
\bibfield{author}{\bibinfo{person}{Jesse Mu} {and} \bibinfo{person}{Advait Sarkar}.} \bibinfo{year}{2019}\natexlab{}.
\newblock \showarticletitle{Do We Need Natural Language? Exploring Restricted Language Interfaces for Complex Domains}. In \bibinfo{booktitle}{\emph{Extended Abstracts of the 2019 CHI Conference on Human Factors in Computing Systems}} (Glasgow, Scotland Uk) \emph{(\bibinfo{series}{CHI EA '19})}. \bibinfo{publisher}{Association for Computing Machinery}, \bibinfo{address}{New York, NY, USA}, \bibinfo{pages}{1–6}.
\newblock
\showISBNx{9781450359719}
\urldef\tempurl%
\url{https://doi.org/10.1145/3290607.3312975}
\showDOI{\tempurl}


\bibitem[Passi and Vorvoreanu(2022)]%
        {passi2022overreliance}
\bibfield{author}{\bibinfo{person}{Samir Passi} {and} \bibinfo{person}{Mihaela Vorvoreanu}.} \bibinfo{year}{2022}\natexlab{}.
\newblock \showarticletitle{Overreliance on AI: literature review}.
\newblock \bibinfo{journal}{\emph{Microsoft Research}} (\bibinfo{year}{2022}).
\newblock


\bibitem[Pirolli and Card(1999)]%
        {pirolli1999information}
\bibfield{author}{\bibinfo{person}{Peter Pirolli} {and} \bibinfo{person}{Stuart Card}.} \bibinfo{year}{1999}\natexlab{}.
\newblock \showarticletitle{Information foraging.}
\newblock \bibinfo{journal}{\emph{Psychological review}} \bibinfo{volume}{106}, \bibinfo{number}{4} (\bibinfo{year}{1999}), \bibinfo{pages}{643}.
\newblock


\bibitem[Pirolli and Card(2005)]%
        {pirolli2005sensemaking}
\bibfield{author}{\bibinfo{person}{Peter Pirolli} {and} \bibinfo{person}{Stuart Card}.} \bibinfo{year}{2005}\natexlab{}.
\newblock \showarticletitle{The sensemaking process and leverage points for analyst technology as identified through cognitive task analysis}. In \bibinfo{booktitle}{\emph{Proceedings of international conference on intelligence analysis}}, Vol.~\bibinfo{volume}{5}. McLean, VA, USA, \bibinfo{pages}{2--4}.
\newblock


\bibitem[Prather et~al\mbox{.}(2023)]%
        {prather2023s}
\bibfield{author}{\bibinfo{person}{James Prather}, \bibinfo{person}{Brent~N Reeves}, \bibinfo{person}{Paul Denny}, \bibinfo{person}{Brett~A Becker}, \bibinfo{person}{Juho Leinonen}, \bibinfo{person}{Andrew Luxton-Reilly}, \bibinfo{person}{Garrett Powell}, \bibinfo{person}{James Finnie-Ansley}, {and} \bibinfo{person}{Eddie~Antonio Santos}.} \bibinfo{year}{2023}\natexlab{}.
\newblock \showarticletitle{" It's Weird That it Knows What I Want": Usability and Interactions with Copilot for Novice Programmers}.
\newblock \bibinfo{journal}{\emph{arXiv preprint arXiv:2304.02491}} (\bibinfo{year}{2023}).
\newblock


\bibitem[Pryzant et~al\mbox{.}(2023)]%
        {Pryzant2023AutomaticPO}
\bibfield{author}{\bibinfo{person}{Reid Pryzant}, \bibinfo{person}{Dan Iter}, \bibinfo{person}{Jerry Li}, \bibinfo{person}{Yin~Tat Lee}, \bibinfo{person}{Chenguang Zhu}, {and} \bibinfo{person}{Michael Zeng}.} \bibinfo{year}{2023}\natexlab{}.
\newblock \showarticletitle{Automatic Prompt Optimization with "Gradient Descent" and Beam Search}. In \bibinfo{booktitle}{\emph{Conference on Empirical Methods in Natural Language Processing}}.
\newblock
\urldef\tempurl%
\url{https://api.semanticscholar.org/CorpusID:258546785}
\showURL{%
\tempurl}


\bibitem[Rothermel et~al\mbox{.}(1998)]%
        {rothermel1998you}
\bibfield{author}{\bibinfo{person}{Gregg Rothermel}, \bibinfo{person}{Lixin Li}, \bibinfo{person}{Christopher DuPuis}, {and} \bibinfo{person}{Margaret Burnett}.} \bibinfo{year}{1998}\natexlab{}.
\newblock \showarticletitle{What you see is what you test: A methodology for testing form-based visual programs}. In \bibinfo{booktitle}{\emph{Proceedings of the 20th international conference on Software engineering}}. IEEE, \bibinfo{pages}{198--207}.
\newblock


\bibitem[Russell et~al\mbox{.}(1993)]%
        {russell1993cost}
\bibfield{author}{\bibinfo{person}{Daniel~M Russell}, \bibinfo{person}{Mark~J Stefik}, \bibinfo{person}{Peter Pirolli}, {and} \bibinfo{person}{Stuart~K Card}.} \bibinfo{year}{1993}\natexlab{}.
\newblock \showarticletitle{The cost structure of sensemaking}. In \bibinfo{booktitle}{\emph{Proceedings of the INTERACT'93 and CHI'93 conference on Human factors in computing systems}}. \bibinfo{pages}{269--276}.
\newblock


\bibitem[Sarkar(2016a)]%
        {sarkar2016constructivist}
\bibfield{author}{\bibinfo{person}{Advait Sarkar}.} \bibinfo{year}{2016}\natexlab{a}.
\newblock \showarticletitle{Constructivist Design for Interactive Machine Learning}. In \bibinfo{booktitle}{\emph{Proceedings of the 2016 CHI Conference Extended Abstracts on Human Factors in Computing Systems}} (San Jose, California, USA) \emph{(\bibinfo{series}{CHI EA '16})}. \bibinfo{publisher}{Association for Computing Machinery}, \bibinfo{address}{New York, NY, USA}, \bibinfo{pages}{1467–1475}.
\newblock
\showISBNx{9781450340823}
\urldef\tempurl%
\url{https://doi.org/10.1145/2851581.2892547}
\showDOI{\tempurl}


\bibitem[Sarkar(2016b)]%
        {sarkar2016phd}
\bibfield{author}{\bibinfo{person}{Advait Sarkar}.} \bibinfo{year}{2016}\natexlab{b}.
\newblock \bibinfo{booktitle}{\emph{{Interactive analytical modelling}}}.
\newblock \bibinfo{type}{{T}echnical {R}eport} UCAM-CL-TR-920. \bibinfo{institution}{University of Cambridge, Computer Laboratory}.
\newblock
\urldef\tempurl%
\url{https://doi.org/10.48456/tr-920}
\showDOI{\tempurl}


\bibitem[Sarkar(2022)]%
        {sarkar2022explainable}
\bibfield{author}{\bibinfo{person}{Advait Sarkar}.} \bibinfo{year}{2022}\natexlab{}.
\newblock \showarticletitle{{Is explainable AI a race against model complexity?}}. In \bibinfo{booktitle}{\emph{{Workshop on Transparency and Explanations in Smart Systems (TeXSS), in conjunction with ACM Intelligent User Interfaces (IUI 2022)}}} \emph{(\bibinfo{series}{CEUR Workshop Proceedings}, \bibinfo{number}{3124})}. \bibinfo{pages}{192--199}.
\newblock
\urldef\tempurl%
\url{http://ceur-ws.org/Vol-3124/paper22.pdf}
\showURL{%
\tempurl}


\bibitem[Sarkar(2023a)]%
        {sarkar2023enough}
\bibfield{author}{\bibinfo{person}{Advait Sarkar}.} \bibinfo{year}{2023}\natexlab{a}.
\newblock \showarticletitle{Enough With “Human-AI Collaboration”}. In \bibinfo{booktitle}{\emph{Extended Abstracts of the 2023 CHI Conference on Human Factors in Computing Systems}} (Hamburg, Germany) \emph{(\bibinfo{series}{CHI EA '23})}. \bibinfo{publisher}{Association for Computing Machinery}, \bibinfo{address}{New York, NY, USA}, Article \bibinfo{articleno}{415}, \bibinfo{numpages}{8}~pages.
\newblock
\showISBNx{9781450394222}
\urldef\tempurl%
\url{https://doi.org/10.1145/3544549.3582735}
\showDOI{\tempurl}


\bibitem[Sarkar(2023b)]%
        {sarkar2023exploring}
\bibfield{author}{\bibinfo{person}{Advait Sarkar}.} \bibinfo{year}{2023}\natexlab{b}.
\newblock \showarticletitle{Exploring Perspectives on the Impact of Artificial Intelligence on the Creativity of Knowledge Work: Beyond Mechanised Plagiarism and Stochastic Parrots}. In \bibinfo{booktitle}{\emph{Proceedings of the 2nd Annual Meeting of the Symposium on Human-Computer Interaction for Work}} (Oldenburg, Germany) \emph{(\bibinfo{series}{CHIWORK '23})}. \bibinfo{publisher}{Association for Computing Machinery}, \bibinfo{address}{New York, NY, USA}, Article \bibinfo{articleno}{13}, \bibinfo{numpages}{17}~pages.
\newblock
\showISBNx{9798400708077}
\urldef\tempurl%
\url{https://doi.org/10.1145/3596671.3597650}
\showDOI{\tempurl}


\bibitem[Sarkar(2023c)]%
        {sarkar2023simplicity}
\bibfield{author}{\bibinfo{person}{Advait Sarkar}.} \bibinfo{year}{2023}\natexlab{c}.
\newblock \showarticletitle{Should Computers Be Easy To Use? Questioning the Doctrine of Simplicity in User Interface Design}. In \bibinfo{booktitle}{\emph{Extended Abstracts of the 2023 CHI Conference on Human Factors in Computing Systems}} (Hamburg, Germany) \emph{(\bibinfo{series}{CHI EA '23})}. \bibinfo{publisher}{Association for Computing Machinery}, \bibinfo{address}{New York, NY, USA}, Article \bibinfo{articleno}{419}, \bibinfo{numpages}{10}~pages.
\newblock
\showISBNx{9781450394222}
\urldef\tempurl%
\url{https://doi.org/10.1145/3544549.3582741}
\showDOI{\tempurl}


\bibitem[Sarkar(2023d)]%
        {sarkar2023eup_genai}
\bibfield{author}{\bibinfo{person}{Advait Sarkar}.} \bibinfo{year}{2023}\natexlab{d}.
\newblock \showarticletitle{Will Code Remain a Relevant User Interface for End-User Programming with Generative AI Models?}. In \bibinfo{booktitle}{\emph{Proceedings of the 2023 ACM SIGPLAN International Symposium on New Ideas, New Paradigms, and Reflections on Programming and Software}} (Cascais, Portugal) \emph{(\bibinfo{series}{Onward! 2023})}. \bibinfo{publisher}{Association for Computing Machinery}, \bibinfo{address}{New York, NY, USA}, \bibinfo{pages}{153–167}.
\newblock
\showISBNx{9798400703881}
\urldef\tempurl%
\url{https://doi.org/10.1145/3622758.3622882}
\showDOI{\tempurl}


\bibitem[Sarkar(2024a)]%
        {sarkar2024challenge}
\bibfield{author}{\bibinfo{person}{Advait Sarkar}.} \bibinfo{year}{2024}\natexlab{a}.
\newblock \showarticletitle{{AI Should Challenge, Not Obey}}.
\newblock \bibinfo{journal}{\emph{Communications of the ACM (in press)}} (\bibinfo{year}{2024}).
\newblock


\bibitem[Sarkar(2024b)]%
        {sarkar2024large}
\bibfield{author}{\bibinfo{person}{Advait Sarkar}.} \bibinfo{year}{2024}\natexlab{b}.
\newblock \showarticletitle{Large Language Models Cannot Explain Themselves}. In \bibinfo{booktitle}{\emph{ACM CHI 2024 Workshop on Human-Centered Explainable AI (HCXAI)}}.
\newblock


\bibitem[Sarkar et~al\mbox{.}(2014)]%
        {sarkar2014teach}
\bibfield{author}{\bibinfo{person}{Advait Sarkar}, \bibinfo{person}{Alan~F Blackwell}, \bibinfo{person}{Mateia Jamnik}, {and} \bibinfo{person}{Martin Spott}.} \bibinfo{year}{2014}\natexlab{}.
\newblock \showarticletitle{Teach and try: A simple interaction technique for exploratory data modelling by end users}. In \bibinfo{booktitle}{\emph{2014 IEEE Symposium on Visual Languages and Human-Centric Computing (VL/HCC)}}. \bibinfo{pages}{53--56}.
\newblock
\urldef\tempurl%
\url{https://doi.org/10.1109/VLHCC.2014.6883022}
\showDOI{\tempurl}


\bibitem[Sarkar et~al\mbox{.}(2020)]%
        {sarkar2020spreadsheet}
\bibfield{author}{\bibinfo{person}{Advait Sarkar}, \bibinfo{person}{Judith~W. Borghouts}, \bibinfo{person}{Anusha Iyer}, \bibinfo{person}{Sneha Khullar}, \bibinfo{person}{Christian Canton}, \bibinfo{person}{Felienne Hermans}, \bibinfo{person}{Andrew~D. Gordon}, {and} \bibinfo{person}{Jack Williams}.} \bibinfo{year}{2020}\natexlab{}.
\newblock \showarticletitle{Spreadsheet Use and Programming Experience: An Exploratory Survey}. In \bibinfo{booktitle}{\emph{Extended Abstracts of the 2020 CHI Conference on Human Factors in Computing Systems}} (Honolulu, HI, USA) \emph{(\bibinfo{series}{CHI EA '20})}. \bibinfo{publisher}{Association for Computing Machinery}, \bibinfo{address}{New York, NY, USA}, \bibinfo{pages}{1–9}.
\newblock
\showISBNx{9781450368193}
\urldef\tempurl%
\url{https://doi.org/10.1145/3334480.3382807}
\showDOI{\tempurl}


\bibitem[Sarkar et~al\mbox{.}(2023)]%
        {sarkar2023participatoryprompting}
\bibfield{author}{\bibinfo{person}{Advait Sarkar}, \bibinfo{person}{Ian Drosos}, \bibinfo{person}{Rob Deline}, \bibinfo{person}{Andrew~D. Gordon}, \bibinfo{person}{Carina Negreanu}, \bibinfo{person}{Sean Rintel}, \bibinfo{person}{Jack Williams}, {and} \bibinfo{person}{Ben Zorn}.} \bibinfo{year}{2023}\natexlab{}.
\newblock \showarticletitle{Participatory prompting: a user-centric research method for eliciting AI assistance opportunities in knowledge workflows}. In \bibinfo{booktitle}{\emph{{Proceedings of the 34th Annual Conference of the Psychology of Programming Interest Group (PPIG 2023)}}}.
\newblock


\bibitem[Sarkar and Gordon(2018)]%
        {sarkar2018spreadsheetlearning}
\bibfield{author}{\bibinfo{person}{Advait Sarkar} {and} \bibinfo{person}{Andrew~D. Gordon}.} \bibinfo{year}{2018}\natexlab{}.
\newblock \showarticletitle{How do people learn to use spreadsheets? (Work in progress)}. In \bibinfo{booktitle}{\emph{{Proceedings of the 29th Annual Conference of the Psychology of Programming Interest Group (PPIG 2018)}}}. \bibinfo{pages}{28--35}.
\newblock


\bibitem[Sarkar et~al\mbox{.}(2022a)]%
        {sarkar2022programmingai}
\bibfield{author}{\bibinfo{person}{Advait Sarkar}, \bibinfo{person}{Andrew~D. Gordon}, \bibinfo{person}{Carina Negreanu}, \bibinfo{person}{Christian Poelitz}, \bibinfo{person}{Sruti Srinivasa~Ragavan}, {and} \bibinfo{person}{Ben Zorn}.} \bibinfo{year}{2022}\natexlab{a}.
\newblock \showarticletitle{What is it like to program with artificial intelligence?}. In \bibinfo{booktitle}{\emph{{Proceedings of the 33rd Annual Conference of the Psychology of Programming Interest Group (PPIG 2022)}}}.
\newblock


\bibitem[Sarkar et~al\mbox{.}(2015)]%
        {sarkar2015interactive}
\bibfield{author}{\bibinfo{person}{Advait Sarkar}, \bibinfo{person}{Mateja Jamnik}, \bibinfo{person}{Alan~F. Blackwell}, {and} \bibinfo{person}{Martin Spott}.} \bibinfo{year}{2015}\natexlab{}.
\newblock \showarticletitle{Interactive visual machine learning in spreadsheets}. In \bibinfo{booktitle}{\emph{2015 IEEE Symposium on Visual Languages and Human-Centric Computing (VL/HCC)}}. \bibinfo{pages}{159--163}.
\newblock
\urldef\tempurl%
\url{https://doi.org/10.1109/VLHCC.2015.7357211}
\showDOI{\tempurl}


\bibitem[Sarkar et~al\mbox{.}(2022b)]%
        {sarkar2022end}
\bibfield{author}{\bibinfo{person}{Advait Sarkar}, \bibinfo{person}{Sruti~Srinivasa Ragavan}, \bibinfo{person}{Jack Williams}, {and} \bibinfo{person}{Andrew~D. Gordon}.} \bibinfo{year}{2022}\natexlab{b}.
\newblock \showarticletitle{End-user encounters with lambda abstraction in spreadsheets: Apollo’s bow or Achilles’ heel?}. In \bibinfo{booktitle}{\emph{2022 IEEE Symposium on Visual Languages and Human-Centric Computing (VL/HCC)}}. \bibinfo{pages}{1--11}.
\newblock
\urldef\tempurl%
\url{https://doi.org/10.1109/VL/HCC53370.2022.9833131}
\showDOI{\tempurl}


\bibitem[Sarkar et~al\mbox{.}(2016)]%
        {sarkar2016visual}
\bibfield{author}{\bibinfo{person}{Advait Sarkar}, \bibinfo{person}{Martin Spott}, \bibinfo{person}{Alan~F. Blackwell}, {and} \bibinfo{person}{Mateja Jamnik}.} \bibinfo{year}{2016}\natexlab{}.
\newblock \showarticletitle{Visual discovery and model-driven explanation of time series patterns}. In \bibinfo{booktitle}{\emph{2016 IEEE Symposium on Visual Languages and Human-Centric Computing (VL/HCC)}}. \bibinfo{pages}{78--86}.
\newblock
\urldef\tempurl%
\url{https://doi.org/10.1109/VLHCC.2016.7739668}
\showDOI{\tempurl}


\bibitem[Schuler and Namioka(1993)]%
        {schuler1993participatory}
\bibfield{author}{\bibinfo{person}{Douglas Schuler} {and} \bibinfo{person}{Aki Namioka}.} \bibinfo{year}{1993}\natexlab{}.
\newblock \bibinfo{booktitle}{\emph{Participatory design: Principles and practices}}.
\newblock \bibinfo{publisher}{CRC Press}.
\newblock


\bibitem[Semnani et~al\mbox{.}(2023)]%
        {Semnani2023WikiChatST}
\bibfield{author}{\bibinfo{person}{Sina~J. Semnani}, \bibinfo{person}{Violet~Z. Yao}, \bibinfo{person}{He Zhang}, {and} \bibinfo{person}{Monica~S. Lam}.} \bibinfo{year}{2023}\natexlab{}.
\newblock \showarticletitle{WikiChat: Stopping the Hallucination of Large Language Model Chatbots by Few-Shot Grounding on Wikipedia}. In \bibinfo{booktitle}{\emph{Conference on Empirical Methods in Natural Language Processing}}.
\newblock
\urldef\tempurl%
\url{https://api.semanticscholar.org/CorpusID:258841157}
\showURL{%
\tempurl}


\bibitem[Serban et~al\mbox{.}(2013)]%
        {serban2013survey}
\bibfield{author}{\bibinfo{person}{Floarea Serban}, \bibinfo{person}{Joaquin Vanschoren}, \bibinfo{person}{J{\"o}rg-Uwe Kietz}, {and} \bibinfo{person}{Abraham Bernstein}.} \bibinfo{year}{2013}\natexlab{}.
\newblock \showarticletitle{A survey of intelligent assistants for data analysis}.
\newblock \bibinfo{journal}{\emph{ACM Computing Surveys (CSUR)}} \bibinfo{volume}{45}, \bibinfo{number}{3} (\bibinfo{year}{2013}), \bibinfo{pages}{1--35}.
\newblock


\bibitem[Shneiderman(1983)]%
        {shneiderman1983direct}
\bibfield{author}{\bibinfo{person}{Ben Shneiderman}.} \bibinfo{year}{1983}\natexlab{}.
\newblock \showarticletitle{Direct manipulation: A step beyond programming languages}.
\newblock \bibinfo{journal}{\emph{Computer}} \bibinfo{volume}{16}, \bibinfo{number}{08} (\bibinfo{year}{1983}), \bibinfo{pages}{57--69}.
\newblock


\bibitem[Siddarth et~al\mbox{.}(2021)]%
        {siddarth2021ai}
\bibfield{author}{\bibinfo{person}{Divya Siddarth}, \bibinfo{person}{Daron Acemoglu}, \bibinfo{person}{Danielle Allen}, \bibinfo{person}{Kate Crawford}, \bibinfo{person}{James Evans}, \bibinfo{person}{Michael Jordan}, {and} \bibinfo{person}{E Weyl}.} \bibinfo{year}{2021}\natexlab{}.
\newblock \showarticletitle{{How AI fails us}}.
\newblock \bibinfo{journal}{\emph{arXiv preprint arXiv:2201.04200}} (\bibinfo{year}{2021}).
\newblock


\bibitem[Srinivasa~Ragavan et~al\mbox{.}(2021)]%
        {srinivasa2021spreadsheet}
\bibfield{author}{\bibinfo{person}{Sruti Srinivasa~Ragavan}, \bibinfo{person}{Advait Sarkar}, {and} \bibinfo{person}{Andrew~D Gordon}.} \bibinfo{year}{2021}\natexlab{}.
\newblock \showarticletitle{Spreadsheet Comprehension: Guesswork, Giving Up and Going Back to the Author}. In \bibinfo{booktitle}{\emph{Proceedings of the 2021 CHI Conference on Human Factors in Computing Systems}} (Yokohama, Japan) \emph{(\bibinfo{series}{CHI '21})}. \bibinfo{publisher}{Association for Computing Machinery}, \bibinfo{address}{New York, NY, USA}, Article \bibinfo{articleno}{181}, \bibinfo{numpages}{21}~pages.
\newblock
\showISBNx{9781450380966}
\urldef\tempurl%
\url{https://doi.org/10.1145/3411764.3445634}
\showDOI{\tempurl}


\bibitem[Stalnaker(2002)]%
        {stalnaker2002common}
\bibfield{author}{\bibinfo{person}{Robert Stalnaker}.} \bibinfo{year}{2002}\natexlab{}.
\newblock \showarticletitle{Common ground}.
\newblock \bibinfo{journal}{\emph{Linguistics and philosophy}} \bibinfo{volume}{25}, \bibinfo{number}{5/6} (\bibinfo{year}{2002}), \bibinfo{pages}{701--721}.
\newblock


\bibitem[Tankelevitch et~al\mbox{.}(2023)]%
        {tankelevitch2023metacognitive}
\bibfield{author}{\bibinfo{person}{Lev Tankelevitch}, \bibinfo{person}{Viktor Kewenig}, \bibinfo{person}{Auste Simkute}, \bibinfo{person}{Ava~Elizabeth Scott}, \bibinfo{person}{Advait Sarkar}, \bibinfo{person}{Abigail Sellen}, {and} \bibinfo{person}{Sean Rintel}.} \bibinfo{year}{2023}\natexlab{}.
\newblock \showarticletitle{The Metacognitive Demands and Opportunities of Generative AI}.
\newblock \bibinfo{journal}{\emph{arXiv preprint arXiv:2312.10893}} (\bibinfo{year}{2023}).
\newblock


\bibitem[Vasconcelos et~al\mbox{.}(2023)]%
        {vasconcelos2023explanations}
\bibfield{author}{\bibinfo{person}{Helena Vasconcelos}, \bibinfo{person}{Matthew J{\"o}rke}, \bibinfo{person}{Madeleine Grunde-McLaughlin}, \bibinfo{person}{Tobias Gerstenberg}, \bibinfo{person}{Michael~S Bernstein}, {and} \bibinfo{person}{Ranjay Krishna}.} \bibinfo{year}{2023}\natexlab{}.
\newblock \showarticletitle{Explanations can reduce overreliance on ai systems during decision-making}.
\newblock \bibinfo{journal}{\emph{Proceedings of the ACM on Human-Computer Interaction}} \bibinfo{volume}{7}, \bibinfo{number}{CSCW1} (\bibinfo{year}{2023}), \bibinfo{pages}{1--38}.
\newblock


\bibitem[Wang et~al\mbox{.}(2021)]%
        {wang2021autods}
\bibfield{author}{\bibinfo{person}{Dakuo Wang}, \bibinfo{person}{Josh Andres}, \bibinfo{person}{Justin~D. Weisz}, \bibinfo{person}{Erick Oduor}, {and} \bibinfo{person}{Casey Dugan}.} \bibinfo{year}{2021}\natexlab{}.
\newblock \showarticletitle{AutoDS: Towards Human-Centered Automation of Data Science}. In \bibinfo{booktitle}{\emph{Proceedings of the 2021 CHI Conference on Human Factors in Computing Systems}} (Yokohama, Japan) \emph{(\bibinfo{series}{CHI '21})}. \bibinfo{publisher}{Association for Computing Machinery}, \bibinfo{address}{New York, NY, USA}, Article \bibinfo{articleno}{79}, \bibinfo{numpages}{12}~pages.
\newblock
\showISBNx{9781450380966}
\urldef\tempurl%
\url{https://doi.org/10.1145/3411764.3445526}
\showDOI{\tempurl}


\bibitem[Wang et~al\mbox{.}(2023)]%
        {Wang2023AssessingTR}
\bibfield{author}{\bibinfo{person}{Weixuan Wang}, \bibinfo{person}{Barry Haddow}, \bibinfo{person}{Alexandra Birch}, {and} \bibinfo{person}{Wei Peng}.} \bibinfo{year}{2023}\natexlab{}.
\newblock \showarticletitle{Assessing the Reliability of Large Language Model Knowledge}.
\newblock \bibinfo{journal}{\emph{ArXiv}}  \bibinfo{volume}{abs/2310.09820} (\bibinfo{year}{2023}).
\newblock
\urldef\tempurl%
\url{https://api.semanticscholar.org/CorpusID:264146357}
\showURL{%
\tempurl}


\bibitem[Wang et~al\mbox{.}(2024)]%
        {Wang2024ChainofTableET}
\bibfield{author}{\bibinfo{person}{Zilong Wang}, \bibinfo{person}{Hao Zhang}, \bibinfo{person}{Chun-Liang Li}, \bibinfo{person}{Julian~Martin Eisenschlos}, \bibinfo{person}{Vincent Perot}, \bibinfo{person}{Zifeng Wang}, \bibinfo{person}{Lesly Miculicich}, \bibinfo{person}{Yasuhisa Fujii}, \bibinfo{person}{Jingbo Shang}, \bibinfo{person}{Chen-Yu Lee}, {and} \bibinfo{person}{Tomas Pfister}.} \bibinfo{year}{2024}\natexlab{}.
\newblock \showarticletitle{Chain-of-Table: Evolving Tables in the Reasoning Chain for Table Understanding}.
\newblock \bibinfo{journal}{\emph{ArXiv}}  \bibinfo{volume}{abs/2401.04398} (\bibinfo{year}{2024}).
\newblock
\urldef\tempurl%
\url{https://api.semanticscholar.org/CorpusID:266899992}
\showURL{%
\tempurl}


\bibitem[Weisz et~al\mbox{.}(2021)]%
        {weisz2021perfection}
\bibfield{author}{\bibinfo{person}{Justin~D Weisz}, \bibinfo{person}{Michael Muller}, \bibinfo{person}{Stephanie Houde}, \bibinfo{person}{John Richards}, \bibinfo{person}{Steven~I Ross}, \bibinfo{person}{Fernando Martinez}, \bibinfo{person}{Mayank Agarwal}, {and} \bibinfo{person}{Kartik Talamadupula}.} \bibinfo{year}{2021}\natexlab{}.
\newblock \showarticletitle{Perfection not required? Human-AI partnerships in code translation}. In \bibinfo{booktitle}{\emph{26th International Conference on Intelligent User Interfaces}}. \bibinfo{pages}{402--412}.
\newblock


\bibitem[Wenskovitch et~al\mbox{.}(2021)]%
        {wenskovitch2021beyond}
\bibfield{author}{\bibinfo{person}{John Wenskovitch}, \bibinfo{person}{Corey Fallon}, \bibinfo{person}{Kate Miller}, {and} \bibinfo{person}{Aritra Dasgupta}.} \bibinfo{year}{2021}\natexlab{}.
\newblock \showarticletitle{Beyond visual analytics: Human-machine teaming for ai-driven data sensemaking}. In \bibinfo{booktitle}{\emph{2021 IEEE Workshop on TRust and EXpertise in Visual Analytics (TREX)}}. IEEE, \bibinfo{pages}{40--44}.
\newblock


\bibitem[Williams et~al\mbox{.}(2020)]%
        {williams2020understanding}
\bibfield{author}{\bibinfo{person}{Jack Williams}, \bibinfo{person}{Carina Negreanu}, \bibinfo{person}{Andrew~D. Gordon}, {and} \bibinfo{person}{Advait Sarkar}.} \bibinfo{year}{2020}\natexlab{}.
\newblock \showarticletitle{Understanding and Inferring Units in Spreadsheets}. In \bibinfo{booktitle}{\emph{2020 IEEE Symposium on Visual Languages and Human-Centric Computing (VL/HCC)}}. \bibinfo{pages}{1--9}.
\newblock
\urldef\tempurl%
\url{https://doi.org/10.1109/VL/HCC50065.2020.9127254}
\showDOI{\tempurl}


\bibitem[Zamfirescu-Pereira et~al\mbox{.}(2023)]%
        {zamfirescu2023johnny}
\bibfield{author}{\bibinfo{person}{JD Zamfirescu-Pereira}, \bibinfo{person}{Richmond~Y Wong}, \bibinfo{person}{Bjoern Hartmann}, {and} \bibinfo{person}{Qian Yang}.} \bibinfo{year}{2023}\natexlab{}.
\newblock \showarticletitle{Why Johnny can’t prompt: how non-AI experts try (and fail) to design LLM prompts}. In \bibinfo{booktitle}{\emph{Proceedings of the 2023 CHI Conference on Human Factors in Computing Systems}}. \bibinfo{pages}{1--21}.
\newblock


\end{thebibliography}
